\documentclass [11pt,useAMS,usenatbib] {article}
\usepackage{amsmath}   
\usepackage{graphicx}   
\usepackage{verbatim}   
\usepackage{color} 
\usepackage{float}     
\usepackage{hyperref}
\usepackage{latexsym}
\usepackage{caption}
\usepackage{soul}
\usepackage[margin=1.0in]{geometry}
\usepackage{natbib}

\begin{document}
\noindent \bf{Evolution of Cosmological Parameters and Fundamental Constants in a Flat Quintessence Cosmology: A Dynamical
Alternative to $\Lambda$CDM}
\vspace{1cm}

\noindent Rodger I. Thompson

\noindent Department of Astronomy and Steward Observatory University of Arizona; rit@arizona.edu

\section{Abstract}
The primary purpose of this work is the provision of accurate, analytic, evolutionary templates for cosmological parameters
and fundamental constants in a dynamical cosmology.  A flat quintessence cosmology with a dark energy potential that
has the mathematical form of the Higgs potential is the specific cosmology and potential addressed in this work.  These
templates, based on the physics of the cosmology and potential are intended to replace the parameterizations currently
used to determine the likelihoods of dynamical cosmologies.  Acknowledging that, unlike $\Lambda$CDM, the evolutions
are dependent on both the specific cosmology and the dark energy potential the templates are referred to as Specific
Cosmology and Potential, SCP, templates.  The requirements set for the SCP templates are that they must be accurate,
analytic functions of an observable such as the scale factor or redshift.  This is achieved through the utilization of a modified
beta function formalism that is based on a physically motivated dark energy potential to calculate the beta function.  The 
methodology developed here is designed to be adaptable to other cosmologies and dark energy potentials. The SCP 
templates are essential tools in determining the relative likelihoods of a range of dynamical cosmologies and potentials. 
An ultimate purpose is the determination whether dark energy is dynamical or static in a quantitative manner.  It is suggested 
that the SCP templates calculated in this work can serve as fiducial dynamical templates in the same manner as 
$\Lambda$CDM serves for static dark energy.

Keywords:\\

Cosmological Parameters ; Fundamental Constants; Quintessence

\section{Introduction} \label{s-int}
This manuscript  examines the evolutions in the late time, matter and dark energy 
dominated, epoch between the scale factors of 0.1 and 1.0 for a flat quintessence cosmology.  This epoch is the primary focus 
of the  upcoming Rubin and Roman observatories observations.  The study does not consider radiation but only
matter and dark energy and is therefore only relevant to late epochs not under the influence of radiation.  The farthest look back
time considered here is at a scale factor of 0.1 where radiation has no measurable effect.

Some preliminary aspects of areas covered in this publication are discussed
in \cite{thm22}.  This work, however, expands the study and is intended for both experts in the field and those who wish an 
introduction to calculations of the evolution of cosmological parameters and fundamental constants. The methodology
presented here is particular to the specific cosmology, quintessence, and the evolutionary templates it calculates are for
a specific dark energy potential.  The templates from the methodology are therefore referred to as Specific Cosmology
and Potential, SCP, templates. 

The nature of dark energy has been declared one of the ``grand challenges in both physics and astronomy'' by the
 Decadal Survey of Astronomy and Astrophysics 2022 \citep{nap21}. A major part of that challenge is the question
 of whether dark energy is static or dynamic.  An important aspect of the question is whether a dynamical cosmology can
 fit the current and future observational data as well or better than $\Lambda$CDM.  This work explores a flat quintessence
 cosmology with current parameter boundary conditions close but not equal to $\Lambda$CDM to provide accurate 
 predictions of parameter and fundamental constant evolutions for comparison with data.  The calculation of the evolutions is 
 described in detail, particularly the use of a modified beta function formalism that 
 produces accurate, analytic, functions of the parameters and constants as a function of the observable scale factor.  The dark 
 energy potential has a natural origin, having the same mathematical polynomial form as the Higgs potential. It is, however, not
 the Higgs field and has none of the rich physics of the Higgs.  It is simply a rolling scalar field with quintessence physics
 that is coupled to gravity.  Since the potential has the same mathematical form as the Higgs potential it is referred to as the
 Higgs Inspired or HI potential. 
 
 The development of the SCP templates for a flat, minimally coupled, Quintessence cosmology takes advantage
 of the property of minimally coupled  systems that the dark energy and matter density evolutions are independent of each other.
 Each can be calculated separately, as is traditionally done for Quintessence. \citep{sch08,cop06,cic17, bah18}, and then combined 
 when necessary for the calculation
 of the cosmological parameters such as the Hubble parameter. As an example the evolution of the matter density is simply
 $\frac{\rho_{m_0}}{a^3}$ where $\rho_{m_0}$ is the current matter density and $a$ is the scale factor.  The evolution of the
 scalar and other dark energy functions are calculated in sections~\ref{s-q} through~\ref{s-hipe} without reference to the matter
 density except in the introduction of the Friedmann constraints.  The matter density is incorporated in section~\ref{s-H} where 
 the Hubble parameter is calculated using the first Friedmann constraint.  An approximation is made in equation~\ref{eq-moden}
 where the $\frac{\beta^2}{6}$ term is set to zero to achieve equation~\ref{eq-qbet} which is only a function of the dark energy
 potential and scalar.  The approximation is valid for all of the dark energy EoS values in this study but could lose accuracy for
 high deviations of $w$ from minus one.
 
Beyond demonstrating the methodology for producing SCP templates for cosmological parameter evolution this work
also examines the role of fundamental constants in setting constraints on both static and dynamical cosmologies. Without
invoking special symmetries it is difficult to prevent a scalar field that couples with gravity from also coupling with other
sectors such as the weak, electromagnetic and strong forces \citep{car98}.  The values of the fundamental constants
such as the fine structure constant $\alpha$ and the proton to electron mass ratio $\mu$ are determined by the Quantum 
Chromodynamic Scale $\Lambda_{QCD}$, the Higgs vacuum expectation value $\nu$ and the Yukawa couplings
$h$ \citep{coc07}. It is assumed here that the HI scalar responsible for dark energy also interacts with these sectors. The
temporal evolution of the constants produced by the interactions is examined in section~\ref{s-fc} along with the connection
to the dark energy Equation of State, EoS, $w$.

The study utilizes  natural units with $\hbar$, $c$ and $8\pi G$ set to one where $G$ is the Newton gravitational constant.  
The mass units are reduced Planck masses $m_p$.  The constant $\kappa$ is the inverse reduced Planck mass
$1/m_p$. In the mass units of this study $\kappa = 1$ but it is retained in equations to display the proper mass units.

\section{The Need for SCP Templates}\label{s-nscpt}
\begin{figure}[H]
\includegraphics[width=10.5 cm]{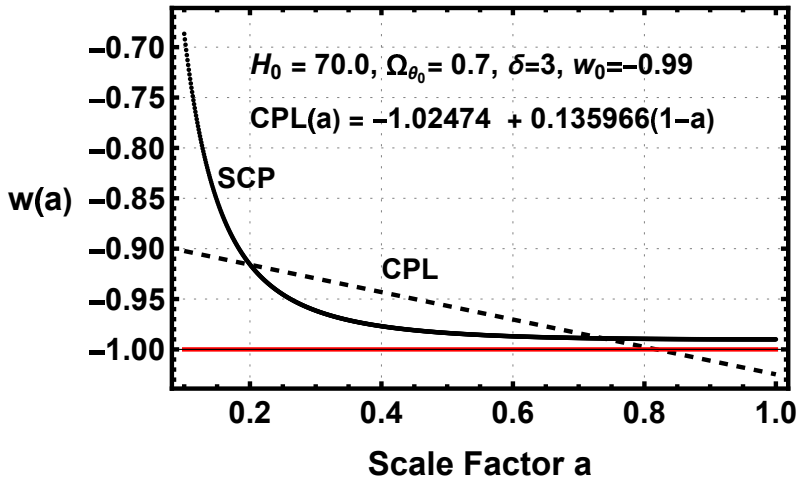}
\caption{The dashed line is the CPL linear fit to the $w(a)$ freezing evolution for $w_0=-0.99$, solid line.}
\label{fig-cpl}
\end{figure}

 The SCP templates generated in this study are candidates for a fiducial set of templates to compare with the 
 observations in the same way that the well known static $\Lambda$CDM templates are currently used.  Although accurate 
 analytic templates for the static $\Lambda$CDM cosmology exist similar templates for dynamical cosmologies are 
 exceedingly rare. Currently the primary tools for analyzing 
dynamical cosmologies are parameterizations such as the Chevallier, Polarski and Linder , CPL, \citep{che01,lin03} 
linear parameterization of the dark energy Equation of State, EoS, $w(a) = w_o+(1-a)w_a$.   Such parameterizations do
not contain any of the physics of the dynamical cosmology and its dark energy potential.   Figure~\ref{fig-cpl} shows the CPL
 fit to a quintessence cosmology $w$ with the dark energy potential described in section~\ref{s-hip}.

The linear CPL, dashed line, is a poor fit to the true evolution, solid line, and is not an accurate measure of its likelihood.   
Beyond producing an erroneous likelihood the CPL fit also produces erroneous conclusions.  At scale factors greater
than 0.8 the CPL fit is in the phantom region, $w<-1$, even though the true evolution has no phantom values. It is well
known that it is quite difficult for quintessence to cross the phantom boundary \citep{vik05}.  Three recent analyses of 
observational data \citep{pla19, div20, div20a} using a MCMC analysis with a CPL dark energy template find phantom 
values of $w$ at low redshift. The presence of phantom values of $w$ can be interpreted as strong evidence against
quintessence, however, figure~\ref{fig-cpl} clearly shows that for a quintessence cosmology, with no phantom values, a
CPL fit erroneously produces phantom values due to fitting a non-linear evolution with a linear parameterization.
The methodology described here produces templates based on the action of the quintessence cosmology and a
specific dark energy potential for comparison to the observational data.  SCP templates are essential to properly 
compare accurate predictions to the observational data in establishing the true likelihood of the cosmology and potential. 

\section{Quintessence}\label{s-q}
As one of the simplest dynamical cosmologies, flat quintessence provides a straightforward example of a methodology
for producing SCP templates.  Quintessence is a well studied \citep{cop06} and well known cosmology but for easy 
reference some important aspects of its physics are given here. 

The quintessence cosmology is defined by its action $S_q$.
\begin{equation}\label{eq-qact}
S_q=\int d^4x\sqrt{-g}\left[m_p^2\frac{R}{2}-\frac{1}{2}g^{\mu\nu}\partial_{\mu}\phi \partial_{\nu}\phi-V(\phi)\right], \hspace{0.5cm}
S_m = \frac{\rho_{m_0}}{a^3}
\end{equation}
where $S_q$ is the dark energy action and $S_m$ is the matter (dust) action. $R$ is the Ricci scalar, $g$ is the 
determinant of the metric $g^{\mu\nu}$, $\phi$ is the scalar, and $V(\phi)$ is the dark energy potential.  The scalar
$\phi$ is the true scalar with a value on the order of $10^{-32} m_p$.  A second scalar $\theta$ is introduced in section~\ref{s-hip}
for the Ratra-Peebles form of the dark energy potential.  This scalar is, therefore, referred to as the Ratra-Peebles, or RP scalar and
has a value on the order of unity in units of $m_p$. In the matter action $\rho_{m_0}$ is the current matter density and $a$ is the scale
factor.  The matter component of the total action $S_m$ is separate from the quintessence
dark energy component $S_q$ resulting in a total action of $S_{tot }= S_q + S_m$ as in equation (3.1) of \citep{cic17} The matter
component is introduced via the first Friedmann constraint in section~\ref{s-H} in the calculation of the Hubble parameter.  In the
following only dark energy is considered in the derivation of the scalar since its evolution is not affected by matter.  Matter is then
introduced in section~\ref{s-H} to derive the proper Hubble parameter that includes both dark energy and matter.

The kinetic component  of $S_q$ is
\begin{equation}\label{eq-X}
X=-\frac{1}{2}g^{\mu\nu}\partial_{\mu}\phi \partial_{\nu}\phi = -\frac{\dot{\phi}^2}{2}.
\end{equation}
The common kinetic symbol $X$ for $-\frac{\dot{\phi}^2}{2}$ is used throughout the manuscript.  $X$ is a function of only 
time since the universe is assumed to be spatially homogeneous.

The quintessence dark energy density and pressure are set by the action as
\begin{equation} \label{eq-rhop}
\rho_{\phi} \equiv -X+V(\phi), \hspace{1cm} p_{\phi} \equiv -X-V(\phi)
\end{equation}
In natural units both the density and the pressure have units of $m^4_p$ and the time derivative of the scalar $\dot{\phi}$ 
has units of $m^2_p$. The dark energy EoS $w(\phi)$ is
\begin{equation} \label{eq-deos}
w(\phi)= \frac{P_{\phi}}{\rho_{\phi}}=\frac{-X-V(\phi)}{-X+V(\phi)}.
\end{equation}
which is a pure number. Combining eqns.~\ref{eq-rhop} gives 
\begin{equation}\label{eq-pdots}
P_{\phi} + \rho_{\phi}  =-2 X
\end{equation}
It follows that
\begin{equation}\label{eq-wpo}
\frac{P_{\phi} +\rho_{\phi}}{\rho_{\phi}} = w+1=\frac{-2 X}{\rho_{\phi}}
\end{equation}
giving $\dot{\phi}$ a relationship to the dark energy EoS and the dark energy density.
\begin{equation} \label{eq-dpw}
-2 X =\rho_{\phi}(w+1).
\end{equation}

\section{General Cosmological Constraints}\label{s-gcc}
Independent of the particular cosmology there are general constraints on the evolution of the cosmological parameters. The
first are the Friedmann constraints.

\subsection{The Friedmann constraints}\label{ss-fcon}
The two Friedmann constraints play an important role in calculating the SCP templates. The forms of the first and second 
Friedman constraints used here are
\begin{equation}\label{eq-friedcs}
3\left(\frac{H(a)}{\kappa}\right)^2 = \rho_{\phi}(a) +\rho_m(a)  \hspace{0.5cm}, \hspace{0.5cm}
3\left(\frac{\dot{H}(a)}{\kappa^2} + \left(\frac{H(a)}{\kappa}\right)^2\right)=-\frac{\rho(a)+3P(a)}{2}
\end{equation}
where $\rho_{\phi}(a)$ is the dark energy density, $\rho_m(a)$ is the matter density, $\rho(a)$ is the sum of the matter and 
dark energy densities and $P(a)$ is the dark energy pressure. In a universe with only dark energy the Friedmann constraints are
\begin{equation}\label{eq-defcon}
3\left(\frac{H_{\phi}(a)}{\kappa}\right)^2=3\left(\frac{\sqrt{\Omega_{\phi}}H(a)}{\kappa}\right)^2 = \rho_{\phi}(a)  \hspace{0.25cm}, 
 \hspace{0.25cm} 3\left(\frac{\dot{H}_{\phi}(a)}{\kappa^2} +\left(\frac{H_{\phi}(a)}{\kappa}\right)^2\right)=-\frac{\rho_{\phi}(a)+3P(a)}{2}
\end{equation}
where $H_{\phi}$ denotes the dark energy only Hubble parameter.  The critical density is $3H^2$ and the ratios of the 
dark energy density and matter density to the critical density are $\Omega_{\phi}$ and $\Omega_m$ that add to one in
a flat universe.  The dark energy only Hubble parameter is then $\sqrt{\Omega_{\phi}}H$.

\subsection{The Boundary Conditions}
Since we are looking for solutions of differential equations the second set of constraints is imposed by the boundary
conditions, the current values of certain cosmological parameters.  Table~\ref{tab-bc} displays the cosmological boundary 
conditions chosen for this study.  The range of the scale factor is slightly arbitrary but is set to include the range covered by 
the Rubin and Roman observations. The range of $w_0$ is set  close to minus one to be near but not exactly 
minus one. The $H_0$ value is set to 73 consistent with the current late time expectations \citep{rie22}.  $\Omega_{m_0}$ and 
$\Omega_{\phi_0}$ are the current concordance values.  All of the boundary conditions appear in the evolutionary functions 
of the SCP templates and are therefore easily changed.
\begin{table}[h]
\caption{ \label{tab-bc} Boundary conditions and parameter values in this work. $H_0$ is the current value of the
Hubble parameter in units of $\frac{km/sec}{Mpc}$. $\Omega_{m_0}$ and $\Omega_{\phi_0}$ are the current ratios 
of the matter and dark energy densities to the critical density. $w_0$ are the current values of the dark
energy equation of state.}
\begin{tabular}{|c|c|c|ccc|c|}
\hline
$H_0$ &$\Omega_{m_0} $&$\Omega_{\phi_0}$& & $ w_0$& &  scale factor $a$\\
\hline
73 & 0.3& 0.7 &-0.99&-0.995&-0.999&0.1 - 1.0\\
\hline
\end{tabular}
\end{table}

\section{The Higgs Inspired Potential}\label{s-hip}
The dark energy potential has the mathematical form of the Higgs potential  $V(\phi)\propto (\phi^2-\gamma^2)^2$ 
a quartic polynomial with a constant $\gamma$. It is chosen for two reasons.  The first is that the mathematical form is 
identical to the Higgs 
potential which gives rise to a scalar field that is known to exist and is therefore physically motivated, hence the name Higgs Inspired or 
HI potential.  A second reason is that by varying the constant term $\gamma$ it produces dark energy equations of state 
that are freezing, thawing, and transitioning between freezing and thawing.  This makes it a good choice for a fiducial potential that covers a wide range of evolutions.

The most convenient form for the potential is a modified Ratra Peebles format \citep{rat88, pee88} with a scalar field denoted 
by $\theta$, the RP scalar introduced in section~\ref{s-q} that has units of mass in $m_p$.  The potential is then
\begin{equation}\label{eq-hirp}
V(\kappa\theta) = M^4((\kappa\theta)^2-(\kappa\delta)^2)^2 =M^4((\kappa\theta)^4-2(\kappa\delta)^2(\kappa\delta)^2 +
(\kappa\delta)^4)
\end{equation}
where the true scalar is $\phi=M\kappa\theta$. $M$ and $\delta$ are constants with units of mass and both $(\kappa\theta)$ 
and $(\kappa\delta)$ are dimensionless therefore all of the dimensionality is in the $M^4$ leading term.  Since both arguments 
are dimensionless there is no need for the $n$ in the usual $M^{4-n}\phi^n$ Ratra Peebles format.  The terms $\theta$ and 
$\delta$ replace the scalar $\phi$ and $\gamma$ terms to differentiate them from the true scalar $\phi$ and constant 
$\gamma$ which have values on the order of $10^{-31}m_p$. The values of $\theta$ and $\delta$ are of the order unity,     
The value of $\delta$ is chosen to be greater than the current scalar $\theta_0$ to place the equilibrium point 
$\theta = \delta$ in the future.  This makes the constant $(\kappa\delta)^4$ the dominant term followed by the two dynamical 
terms $-2(\kappa\theta)^2(\kappa\delta)^2$ and $(\kappa\theta)^4$ in descending order.

\section{The Quintessence Methodology}\label{s-qm}
The methodology developed here is for the specific quintessence cosmology and only applies to a cosmology whose action
is given by equation~\ref{eq-qact}.  Among the current plethora of dynamical cosmologies there are some with quite different 
names that have the same action as quintessence where this methodology will apply but not for cosmologies such as
k-essence that has a different action. The methodology is demonstrated with the modified Ratra Peebles HI potential of 
equation~\ref{eq-hirp}.

\subsection{The Modified Beta Function Formalism}\label{ss-bff}
The key to producing SCP templates that are accurate analytic functions of the scale factor is the beta function formalism
\citep{bin15, bin17, cic17, koh17}.  The beta function is a differential equation relating the scalar $\kappa\theta$ to the scale 
factor $a$ allowing the calculation of an analytic form of $\kappa\theta(a)$. The analytic form for the scalar is
achieved via the approximation of a dominant potential component of the dark energy density that allows the exclusion of the
kinetic component to calculate the beta function given in equation~\ref{eq-qbet}.  This, in turn, enables the calculation of the
analytic function for the scalar.  The cosmological parameters are analytic functions of the scalar that are quite accurate but
not exact.  The cosmological parameter templates do not contain any numerical calculations.

The primary beta function formalism papers 
relative to this work are \citep{bin15, cic17}.  The work by \citep{bin15} considers a quintessence dark energy only universe
while the work of \citep{cic17} considers a quintessence universe with both matter and dark energy which is the universe
considered in this work.  Both \citep{bin15} and \citep{cic17} consider the general physics of the beta function formalism
rather than the explicit evolution of cosmological parameters.  Their approach is therefore modified in this work to provide
analytic evolutionary templates for cosmological parameters.  These modifications are noted in the following discussion.

The generalized beta function \citep{bin17} is defined as
\begin{equation}\label{eq-gbeta}
\beta(\kappa\theta) \equiv \left (-\frac{\partial p}{\partial X}\right)^{\frac{1}{2}}\frac{d\kappa\theta}{d \ln(a)}
\end{equation}  
From equation.~\ref{eq-rhop} for the quintessence dark energy pressure it is evident that $(-\frac{\partial p}{\partial X})^{\frac{1}{2}}$ 
is one, giving a quintessence beta function of 
\begin{equation} \label{eq-qbeta}
\beta(\kappa\theta) \equiv \frac{d\kappa\theta}{d \ln(a)}=\frac{d\kappa\theta}{da}a.
\end{equation}
From equation~\ref{eq-qbeta} and the definition of the quintessence beta function and the Hubble parameter
\begin{equation}\label{eq-bh}
\frac{d\kappa\theta}{da} = \frac{\beta}{a}, \hspace{0.5cm}  \kappa\dot{\theta} = \beta H_{\theta}.
\end{equation}
The dark energy only Hubble parameter $H_{\phi}$ is used in equation~\ref{eq-bh} to be consistent with the
dark energy only derivation of the scalar, however, when the matter density is introduced in section~\ref{s-H} the Hubble parameter
for both dark energy and matter should be used since it sets the time evolution of the scale factor $\frac{da}{dt}$.
Equations~\ref{eq-dpw} and~\ref{eq-bh} provide the useful relation
\begin{equation}\label{eq-bow}
\beta(\kappa\theta)=\sqrt{3\Omega_{\theta}(w+1)}
\end{equation}

Since the beta function formalism is developed
for dark energy the first Friedmann constraint in equation~\ref{eq-defcon} applies and
\begin{equation}\label{eq-hrho}
3H_{\theta}^2=\rho_{\theta} = -X +V(\theta)= \frac{(\beta H_{\theta})^2}{2} + V(\theta)
\end{equation}
where now the subscript $\theta$ designates the dark energy density.

In \citep{bin15, cic17} the beta function is defined as the negative of the logarithmic derivative of the dark energy density.
To achieve the desired analytic SCP templates equation~\ref{eq-hrho} is rearranged to a slightly modified density
\begin{equation}\label{eq-moden}
3H_{\theta}^2 \left(1-\frac{\beta^2}{6}\right)= V(\theta)
\end{equation}
noting that $\frac{\beta^2}{6} \ll 1$ for all of the cases considered here .  Using the modified density the beta 
function is then the negative of the logarithmic derivative of the analytic HI potential.
\begin{equation}\label{eq-qbet}
\beta(\kappa\theta)=- \frac{ \frac{\partial V(\kappa\theta)}{\partial(\kappa\theta)}}{V(\kappa\theta)}=
\frac{\partial(\kappa\theta)}{\partial \ln a}.
\end{equation}
The leading constant, $M^4$, in the dark energy potential does not appear in the logarithmic derivative defining the
beta function leaving it as an adjustable parameter to satisfy the Friedmann constraints.   
Note that the approximation that $\frac{\beta^2}{6} \ll 1$ is equivalent to the statement that the kinetic term
$X =- \frac{(\beta H_{\theta})^2}{2}$ is small compared to the dark energy potential.  This is roughly equivalent to the slow roll condition
often used in evaluating dynamical cosmologies.  In fact equation~\ref{eq-qbet} is the negative of the first slow roll condition.  The
first slow roll condition is often set to a small constant, eg. \citep{sch08}, which is only valid for an exponential potential. Here the 
approximation of a small value of $X$ is only used to calculate the analytic form of the scalar and the non-constant time derivative 
of the scalar is used in all parameter calculations. The approximation that $\beta H_{\theta}$ is set to zero in determining the beta
function also means that the Hubble parameter is not used in the derivation of the scalar and that its value of either $H_{\theta}$ 
or $H$ is not a factor.

Although the beta density $3H^2(1-\frac{\beta^2}{6})$ is slightly different than the real density, application of the 
boundary conditions and the Friedmann constraints produces evolutionary SCP templates of high accuracy as
illustrated in section~\ref{s-ac}.

\subsubsection{The beta function for the HI potential}\label{sss-bhi}
Using the HI potential in equation~\ref{eq-hirp} the negative of the logarithmic derivative is
\begin{equation}\label{eq-bhi}
\beta(\kappa\theta) = \frac{-4\kappa\theta}{(\kappa\theta)^2-(\kappa\delta)^2}=\sqrt{3\Omega_{\theta}(w+1)}
\end{equation}
where the last term is from equation~\ref{eq-bow}.  Solving the equation formed by the last two terms for the current time
yields the current value $\theta_0$ for the scalar which is an important boundary condition.
\begin{equation}\label{eq-kto}
\kappa\theta_0=-\frac{4-\sqrt{16+12\Omega_{\theta_0}(w_0+1)(\kappa\delta)^2}}{2\sqrt{3\Omega_{\theta_0}(w_0+1)}}
\end{equation}
Since the argument of the square root in the numerator is greater than 16 equation~\ref{eq-kto} is the positive solution of the 
quadratic equation.

\subsection{The scalar as a function of the scale factor}\label{ss-tofa}
An essential step in achieving SCP templates as analytic functions of the scale factor $a$ is finding the scalar $\kappa\theta$
as a function of $a$.  From the definition of the beta function in equation~\ref{eq-qbet} the differential equation for the scalar as a
function of the scale factor is
\begin{equation}\label{eq-tdifq}
\frac{\partial(\kappa\theta)}{\partial \ln a}= \frac{-4\kappa\theta}{(\kappa\theta)^2-(\kappa\delta)^2}.
\end{equation}
Separating the scale factor and scalar terms gives
\begin{equation}\label{eq-tb}
4 d(\ln(a))=-\frac{(\kappa\theta)^2-(\kappa\delta)^2}{\kappa\theta}d(\kappa\theta).
\end{equation}
An integral of both sides of equation~\ref{eq-tb} from 1 to $a$ for the left side and from $\theta_0$ to $\theta$ on the right side
gives
\begin{equation}\label{eq-intb}
8\ln(a)=2 (\kappa\delta)^2\ln(\kappa\theta)-(\kappa\theta)^2-\left(2 (\kappa\delta)^2\ln(\kappa\theta_0)-(\kappa\theta_0)^2\right).
\end{equation}
Equation~\ref{eq-kto} provides the value of $\kappa\theta_0$.
The following manipulations from \citep{thm22} provide a solution for $\kappa\theta(a)$ involving the Lambert W
function in terms of a constant $c=2 (\kappa\delta)^2\ln(\kappa\theta_0)-(\kappa\theta_0)^2$ and the scale factor.
Dividing both sides of equation~\ref{eq-intb} by $(\kappa\delta)^2$ gives
\begin {equation}\label{eq-bg}
\frac{8}{(\kappa\delta)^2}\ln(a)+\frac{c}{(\kappa\delta)^2}=2\ln(\kappa\theta)-\left(\frac{\theta}{\delta}\right)^2
\end{equation}
Taking the exponential of both sides of equation~\ref{eq-bg} and dividing again by $(\kappa\delta)^2$ yields
\begin{equation}\label{eq-ta}
-\frac{a^{\frac{8}{(\kappa\delta)^2}}}{(\kappa\delta)^2}e^{\frac{c}{(\kappa\delta)^2}}=
-\left(\frac{\theta}{\delta}\right)^2e^{-\left(\frac{\theta}{\delta}\right)^2}
\end{equation}
Equation~\ref{eq-ta} has the mathematical form of the Lambert W function that is the solution to
\begin{equation}\label{eq-W}
\chi= W(\chi)e^{W(\chi)}
\end{equation}
where
\begin{equation}\label{eq-wx}
\chi(a)=-\frac{a^{\frac{8}{(\kappa\delta)^2}}}{(\kappa\delta)^2}e^{\frac{c}{(\kappa\delta)^2}} \hspace{1cm} W(\chi) = -\left(\frac{\theta}{\delta}\right)^2.
\end{equation}
Equations~\ref{eq-W} and~\ref{eq-wx} provide an analytic solution for $\kappa\theta)a)$
\begin{equation}\label{eq-twa}
\kappa\theta(a)=\kappa\delta\sqrt{-W \left(-\frac{a^{\frac{8}{(\kappa\delta)^2}}}{(\kappa\delta)^2}e^{\frac{c}{(\kappa\delta)^2}}\right)}
\end{equation}
that is the positive solution for the square root which is real since $W(x)$ is negative.  The following variable changes
produce a concise form for $\kappa\theta(a)$.
\begin{equation}\label{eq-var}
q=-\frac{e^{\frac{c}{(\kappa\delta)^2}}}{(\kappa\delta)^2} \hspace{0.15cm}, \hspace{0.15cm} p =\frac{8}{(\kappa\delta)^2}
\hspace{0.15cm}, \hspace{0.15cm}\chi(a) = q a^p
\end{equation} 
which yields
\begin{equation}\label{eq-thchi}
\kappa\theta(a) = \kappa\delta \sqrt{-W(\chi(a))}
\end{equation}
Equation~\ref{eq-thchi} provides the key to transforming evolutions that are a function of the scalar into functions of the
observable scale factor to produce the SCP templates.

The term $(\kappa\theta)^2-(\kappa\delta)^2$ appears often in this manuscript.  In terms of the Lambert W function 
$W(\chi(a))$ it is
\begin{equation}\label{eq-tdw}
(\kappa\theta)^2-(\kappa\delta)^2 = -(\kappa\delta)^2(W(\chi(a)) + 1).
\end{equation}
The form of the HI potential is then
\begin{equation}\label{eq-depw}
V(a)=(M\kappa\delta)^4(W(\chi(a)) + 1)^2.
\end{equation}
The beta function also has a compact form in the W function format
\begin{equation}\label{eq-bfw}
\beta(\kappa\theta)= \frac{-4\kappa\theta}{(\kappa\theta)^2-(\kappa\delta)^2}=\frac{4\sqrt{-W(\chi(a))}}{\kappa\delta(W(\chi(a)) +1)}.
\end{equation}

\subsection{Summary of the methodology}\label{ss-sm}
Although the methodology may appear to be complex the separate steps of the quintessence beta function formalism are relatively
simple.  The quintessence beta function is given in equation~\ref{eq-qbeta} which connects the scalar $\kappa\theta$ to the scale
factor $a$. As stated in the text the beta function is defined as the negative of the logarithmic derivative of the dark energy density
but here it is noted that the kinetic term in the density is small compared to the potential term and a legitimate approximation is
to ignore it and set the beta function to the negative of the logarithmic derivative of the potential as given in equation~\ref{eq-qbet}.
This is the only place in the methodology that $\frac{\beta^2}{6}$ is set to zero.  The kinetic term $-X$, as shown in equation~\ref{eq-hrho},
is used in all subsequent calculations of the templates.   Equation~\ref{eq-bhi} shows the beta function calculated from 
equation~\ref{eq-qbet} and is shown in differential form in equations~\ref{eq-tdifq} and~\ref{eq-tb}.  The solution for $\kappa\theta(a)$
is achieved through mathematical manipulation to the simple form in equation~\ref{eq-thchi}. This provides a solution for
a cosmological parameter as a function of the scale factor if the solution as a function of the scalar is known.

Since the matter (dust) action is separate from the dark energy action the evolution dark energy scalar is calculated from 
the dark energy action as described above.  The matter is included in section~\ref{s-H} that derives the Hubble parameter which is a 
function of the dark energy and matter.  It is added to the total density in equation~\ref{eq-hfc1} for the first Friedmann constraint and
is present in the Hubble parameter template in equation~\ref{eq-H}.  This is the Hubble parameter that is used in the calculation of the
time derivative of the scalar $\dot{\theta} =\beta H$.  Adherence to the Friedmann constraints and the inclusion of the matter density
in the Hubble parameter that calculates $\dot{\theta}$  produces accurate SCP templates that are functions of both the matter and the 
dark energy densities and conforms to both Friedmann constraints. 

\section{The Cosmology of $W(\chi)$}\label{s-cw}
\begin{figure}[H]
\includegraphics[width=14.0 cm]{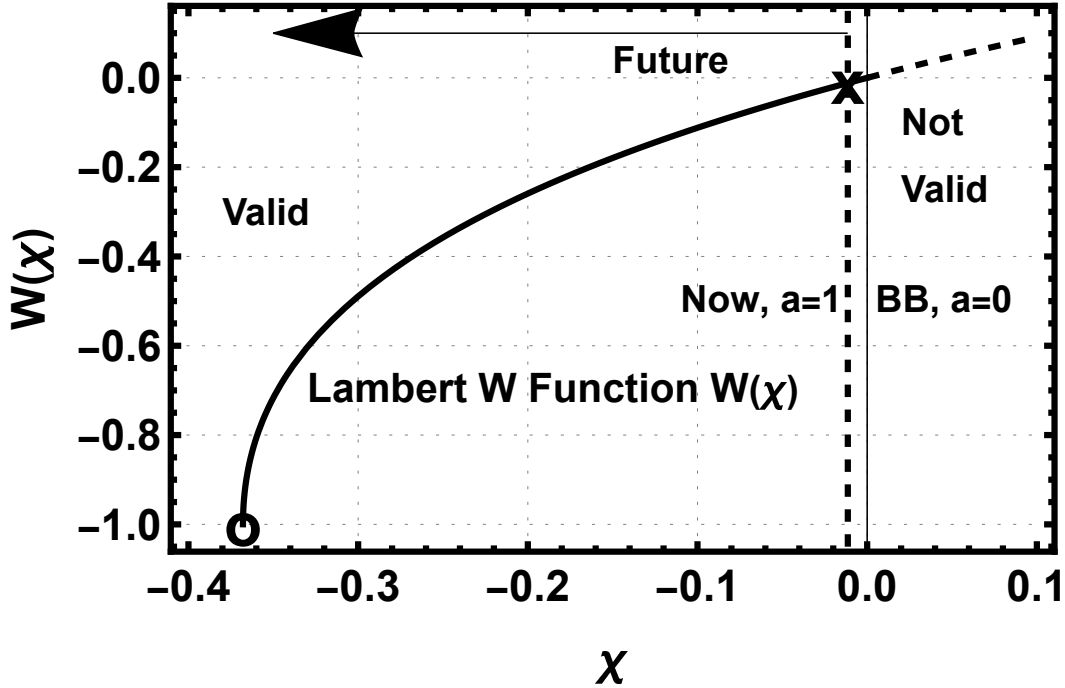}
\caption{The figure show the valid region with negative $\chi$ and invalid regions with positive $\chi$ of the W function 
for this work  The thin vertical line at $\chi=0$, $a=0$, is the big bang.  The X on the track is the present day 
location and the O at the end of the track is when the acceleration goes to zero.  A more detailed description of the figure is 
given in the text.}
\label{fig-chil}
\end{figure}

Before moving on to consider the evolution of $\kappa\theta(a)$, $\beta(a)$ and other parameters it is worthwhile to examine 
the cosmology embedded in the evolution of $W(\chi)$. A thorough discussion of the 
mathematical properties of the Lambert W function is in \citep{olv10}. There it is shown that $W$ has negative values if
its argument is between $-\frac{1}{e}$ and zero which is true for all cases considered here. This makes the argument of
the square root in equation.~\ref{eq-thchi} positive producing a real value of the scalar $\kappa\theta$.  Figure~\ref{fig-chil} 
shows the evolution of the principal branch of $W(\chi)$. The formal designation of the  principal
branch is $W_0(\chi)$ but the subscript is dropped in the following since only the principal branch is used in this work.
Figure~\ref{fig-chil} shows the evolution of $W(\chi$) as a solid line for negative $\chi$ and a dashed line for positive $\chi$.  The
negative portion of the Lambert W function terminates at $\chi= -\frac{1}{e}$ while the positive portion continues indefinitely.  
Only the negative portion has real solutions for the scalar.  Equation~\ref{eq-wx} shows that $\chi$ is negative for all
positive values of the scale factor $a$.  Evolution in figure~\ref{fig-chil} proceeds from right to left as the top arrow indicates.
The variable $\chi$ is zero when $a=0$ therefore the big bang is at $\chi=0$ shown by the vertical thin line.  The dashed vertical 
line just to the left of the big bang shows the maximum excursion of the greatest evolution case, $\delta=3$ and $w_0=-0.99$.
All of the cases considered here have $-W(\chi)$ values much less than one which means that $\theta \ll \delta$. 

\begin{figure}[H]
\includegraphics[width=14.0 cm]{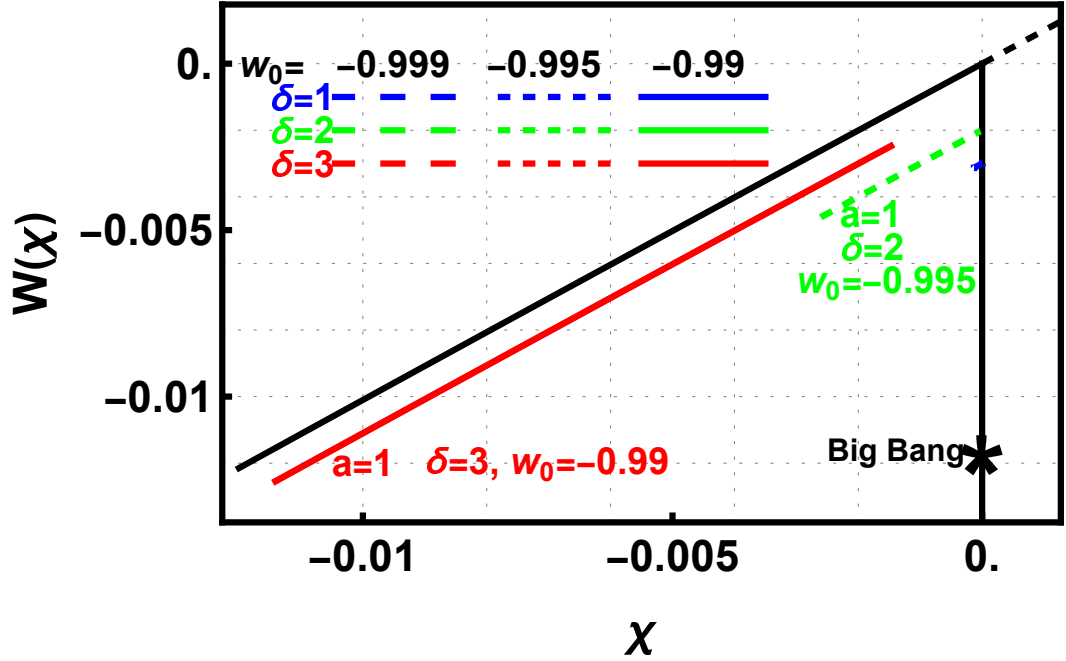}
\caption{The figure shows the region between the big bang and the furthest evolution of any of the cases in this study. This
figure initiates a code continued throughout the paper. The $\delta$ = 1, 2 and three cases are displayed in red, green and
blue. The $w_0$ = -0.99, -0.995 and-0.999 cases are displayed with solid, dashed and long dashed line styles
as shown in the upper left of the figure. The figure is an expanded view of the region between the thin black vertical line and the 
thick dashed line in Figure~\ref{fig-chil}. In this figure the big bang is the black vertical line marked with an asterisk near the bottom.  
The right and left ends of the evolutions are the start point at $a=0.1$ and the end point at $a=1.0$ respectively.  Further discussion
of the figure is in the text.}
\label{fig-chis}
\end{figure}

Figure~\ref{fig-chis} shows the detail of the evolution region of figure~\ref{fig-chil}. The black solid and dashed lines are the same 
as in figure~\ref{fig-chil} but only for the evolution region between the thick dashed and thin solid vertical lines. The 
first case, $\delta=3$ $w_0=-0.99$ in red, with the highest $\delta$ and largest deviation of $w_0$ from minus one has the
most extensive evolution with the right end of the evolution, $a=0.1$, significantly after the big bang.  Note that the evolutions
which actually overlap the black lines have been shifted downward for visibility. The third case with the lowest value of $\delta$
and the least deviation of $w_0$ from minus one has the least evolution with its $a=0.1$ start $\chi$ at $-1.312\times 10^{-12}$
and its maximum $\chi$ of $1.3119\times10^{-4}$ barely visible on the diagram. The middle excursion with $\delta$ = 2 and
$w_0$ = -0.995 starts with $\chi$ = $2.604\times10^{-5}$ and a current $\chi$ of  $2.6404\times10^{-3}$.  The similar numbers
are due to the end scale factor being ten times the beginning scale factor.  Recall that $\chi$ is not time so that a small value of 
$\chi$ does not mean that the scale factor of 0.1 is very near the big bang.  As hinted at in figure~\ref{fig-chil} the evolution region's
small extent make the evolutionary tracks in $\chi$ to appear linear.  Figure~\ref{fig-chis} is an indication, verified later on,
that the value of $\delta$ has a strong influence on the SCP templates. The upper left of the figure shows the 
color code of the $\delta$ values and the line styles of the $w_0$ values maintained through out the manuscript.

The black sloped solid and dashed line is the same as in Figure~\ref{fig-chil} for the small expanded region.  The
red solid and green dashed line are the evolutions of the $\delta=3$, $w_0=-0.99$ and the $\delta=2$ $w_0=0.995$ cases.
The long dashed blue evolution for the $\delta=1$ $w_0=-0.999$ is so short that it appears only as a small dot near $\chi=0$
below the $\delta= 3$ and 2 evolutions. The evolutions are offset downward from the black track for visibility.

To aid comprehension of Figure~\ref{fig-chis} table~\ref{tab-chi} gives the value of $\chi$ at scale factors of 01.,
0.5 and 1.0 for all values of $\delta$ and $w_0$. All values of $\chi$ are negative  The magnitude of $\chi$ increases with increasing
values of $\delta$ and higher deviations of $w_0$ from minus one.  The time evolution of the tracks in Figure~\ref{fig-chis} is
from right to left, the same as in Figure~\ref{fig-chil}. The time extents for all tracks are the same, $a=0.1$ to 1.0, but the 
evolution of $\chi$ has a large variation.

\begin{table}[h]
\caption{ \label{tab-chi}The values of $\chi$ for all values of $\delta$ and $w_0$ for scale factors of 0.1, 0.5 and 1.0.  The barely
visible Figure~\ref{fig-chis} blue $\delta=1.$ and $w_0=-0.999$ $\chi$ values are given by the last row of the $\delta=1$ $\chi$
values in the table. The equilibrium values of the scale factor $a_{eq}$, discussed in section~\ref{ss-fe}, are in the last column.} 
\begin{tabular}{|c|c|c|c|c|c|}
\hline
& &\multicolumn{3}{|c|}{The Values of $\chi$} &\\
\hline
 & &\multicolumn{3}{|c|}{ scale factor $a$}& \\
\hline
$\delta$ & $w_0$ .& 0.1 &0.5 &1.0 & $a_{eq}$ \\
 \hline
 1. & -0.99&  $ -1.307 \times 10^{-11}$ & $-5.107 \times 10^{-6}$ &  $-0.00131$&2.024\\
\hline
 1. & -0.995&  $ -6.550 \times 10^{-12}$ & $-2.558 \times 10^{-6}$ &  $-0.000655$&2.206\\
\hline
 1. & -0.999&  $ -1.312 \times 10^{-12}$ & $-5.125 \times 10^{-7}$ &  $-0.000131$&2.698\\
\hline
\hline
 2. & -0.99&  $ -0.0000517$ & $-0.00129$ &  $-0.00517$&8.437\\
\hline
 2. & -0.995&  $ -0.0000260$ & $-0.000651$ & $-0.00260$&11.885\\
\hline
 2. & -0.999&  $ -5.243 \times 10^{-6}$ & $-0.000131$ & $-0.000524$&26.492\\
\hline
\hline
 3. & -0.99&  $ -0.00147$ & $-0.00616$ &  $-0.0114$&49.776\\
\hline
 3. & -0.995&  $ -0.000750$ & $-0.00313$ & $-0.00580$&106.479\\
\hline
 3. & -0.999&  $ -0.000152$ & $-0.000636$ & $-0.00118$&640.859\\
\hline
\end{tabular}
\end{table}

\subsection{Past Evolution}\label{ss-pe}
The main content of this manuscript is the past evolution from the present to a past scale factor of 0.1 which is a redshift of
nine.  This encompasses a large fraction of the history of the universe in the matter and dark energy dominant epochs. An 
important question is how far back can the SCP templates be utilized.  A hard limit is the onset of the radiation dominated
epoch since the radiation density is not included in the present work.  A reasonable limit is when the radiation density is $1\%$
of the matter density.  The present matter density for $H_0=73$ and $\Omega{m_0}$ of 0.3 is $3.68\times10^{-121}$ 
$m_p^4$ and a present radiation density of $6.17\times10^{-125}$ $m_p^4$. The radiation density is $1\%$ of the matter 
density at a scale factor of 0.0168 or a redshift of 58.5.  This is strictly a physics limit on the validity of the templates. 
Figures~\ref{fig-chil} and~\ref{fig-chis} plus table~\ref{tab-chi} indicate that the template for the scalar $\theta$ is valid back
to this limit but the templates have not been tested for mathematical stability at scale factors smaller than 0.1.  Inclusion of
the radiation density is beyond the scope of this work but it can probably be included in the same manner as the matter density.

\subsection{Future Evolution}\label{ss-fe}
A perhaps even more intriguing question is how far in the future can the templates be extended.  The solution of $\theta$ is
analytic at scale factors greater than one which is all of the region to the left of the dashed vertical line in Figure~\ref{fig-chil}.
There is a limit however to the principal branch of the Lambert W function at $\chi=-\frac{1}{e}$. Figure~\ref{fig-chil} marks  
this location with a O at $W(\chi) = -1$.  Equation~\ref{eq-thchi} shows that at $W(\chi) = -1$ $\theta=\delta$ which is an 
equilibrium point where the dark energy potential is zero.  The scale factor, $a_{eq}$, where this occurs is given by
\begin{equation}\label{eq-aeq}
a_{eq} = (\frac{-1}{q e})^{\frac{1}{p}}.
\end{equation}
where $p$ and $q$ are the same as in equation~\ref{eq-var}. The values of $a_{eq}$ are listed in the last column of Table~\ref{tab-chi}.
It is beyond the scope of this manuscript to determine whether this is a stable equilibrium point. If it is a stable equilibrium, with
$\dot{\theta}$ also zero, then it would be the end of dark energy acceleration. The universe would return to a matter dominated 
evolution. with $3H^2=\frac{\rho_{m_0}}{a^3}$ making a graceful exit from acceleration.  The speed of expansion would be
\begin{equation}\label{eq-madot}
\dot{a} = a H =\sqrt{\frac{\rho_{m_0}}{a}}.
\end{equation}
The universe would then evolve in a classical manner, slowing down to zero expansion at infinity. Given the past history of the universe 
it is reasonable that the lowering level of total density might reveal a new source of accelerated expansion whose  density is below that 
of the current density.

\section{The Evolution of the Scalar and the Beta Function}\label{s-esb}
The scalar $\kappa\theta(a)$ and the beta function $\beta(a)$ influence the evolution of all of the cosmological parameters
in this study.  The sections below document their evolution.

\subsection{The Evolution of $\kappa\theta(a)$}\label{ss-es}
\begin{figure}[H]
\includegraphics[width=14.0 cm]{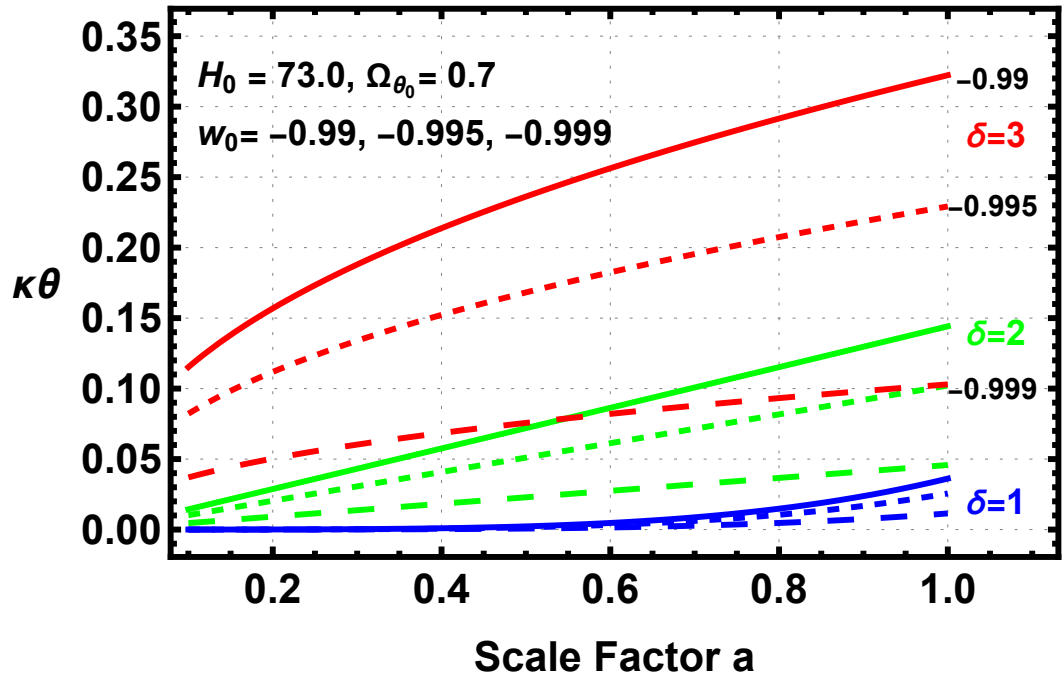}
\caption{The figure plots the evolution of the scalar for all of the $\delta$ and $w_0$ values in this study.}
\label{fig-es}
\end{figure}

Figure~\ref{fig-es} shows the evolution of the scalar $\kappa\theta$ over the scale factors considered in this work.  The colors
and line styles are consistent with the previous figure.
The $w_0$ values for the $\delta=3$ case are labeled on the plot. The order and line styles are the same for the other two 
$\delta$ values.  The evolution is relatively small consistent with a slow roll. As expected the scalar values are
monotonically increasing.  The most striking feature is that the second derivative of the evolution changes from positive for
$\delta=1$ to almost zero for $\delta=2$ to negative for $\delta=3$.  The three, seemingly arbitrary, delta values were chosen
to illustrate this transition.  The transition has a large effect on some parameters, such as the dark energy EoS, $w$, but
relatively little effect on the Hubble parameter as is shown later in Figure~\ref{fig-eh}.

\subsection{The Evolution of $\beta(a)$}\label{ss-eb}
\begin{figure}[H]
\includegraphics[width=14.0 cm]{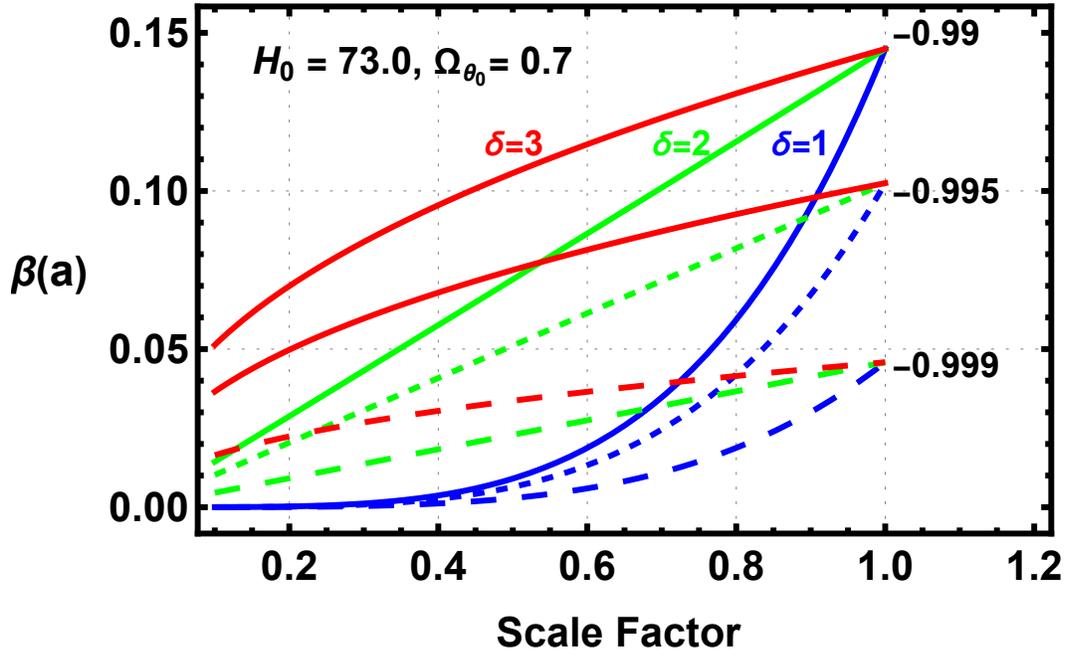}
\caption{The figure shows the evolution of the beta function for all cases in this study.}
\label{fig-ebf}
\end{figure}

The beta function evolution is shown in figure~\ref{fig-ebf}
Although the general nature of the evolution of the beta function is different from the scalar it shows the same change in
the second derivative of the evolution, positive for $\delta=1$, almost zero for $\delta=2$ and negative for $\delta=3$.
The absolute value of $\beta$ is small and decreases as $w_0$ approaches minus one.  The current value of beta,
$\beta_0$ is identical for a given value of $w_0$ due to equation~\ref{eq-bow} which sets $\beta_0$ at $\sqrt{3
\Omega_{\theta_0}(w_0+1)}$ where the subscript 0 indicates the current values.  The beta function appears in
many cosmological parameters due to equation~\ref{eq-bh} that links $\kappa\dot{\theta}$ and the Hubble parameter.

\section{The Value of $M$ in the Dark Energy Potential}\label{s-vm}
At this point the value of $M$ in equation~\ref{eq-hirp} has not been calculated.  The Friedmann constraints and 
equation.~\ref{eq-rhop} provide the means of calculating $M$.  The dark energy density is
\begin{equation}\label{eq-denm}
\rho_{\theta} = \frac{(\kappa\dot{\theta})^2}{2} +M^4((\kappa\theta)^2-(\kappa\delta)^2)^2 = 3\Omega_{\theta}H^2.
\end{equation}
Using $\kappa\dot{\theta} = \beta H$
\begin{equation}\label{eq-3h}
3\Omega_{\theta}H^2 = \frac{(\beta H)^2}{2} + M^4((\kappa\theta)^2 -(\kappa\delta)^2)^2.
\end{equation}
Since $M$ is a constant it can be set using the current boundary conditions which insures adherence to
the first Friedmann constraint at a scale factor of one.  This eliminates any constant offsets due to the approximation in
equation~\ref{eq-qbet} further improving the accuracy of the templates.

Rearranging equation~\ref{eq-3h} and using the current values of the parameters gives
\begin{equation}\label{eq-mp}
3H_0^2(\Omega_{\theta_0}-\frac{\beta_0^2}{6}) = M^4((\kappa\theta_0)^2-(\kappa\delta)^2)^2, \hspace{0.5cm}M=\sqrt[4]{
{\frac{3H_0^2}{((\kappa\theta_0)^2-(\kappa\delta)^2)^2}\left(\Omega_{\theta_0} -\frac{\beta_0^2}{6}\right)}}.
\end{equation}

\section{The Evolution of the HI Dark Energy Potential}\label{s-hipe}
\begin{figure}[H]
\includegraphics[width=14.0 cm]{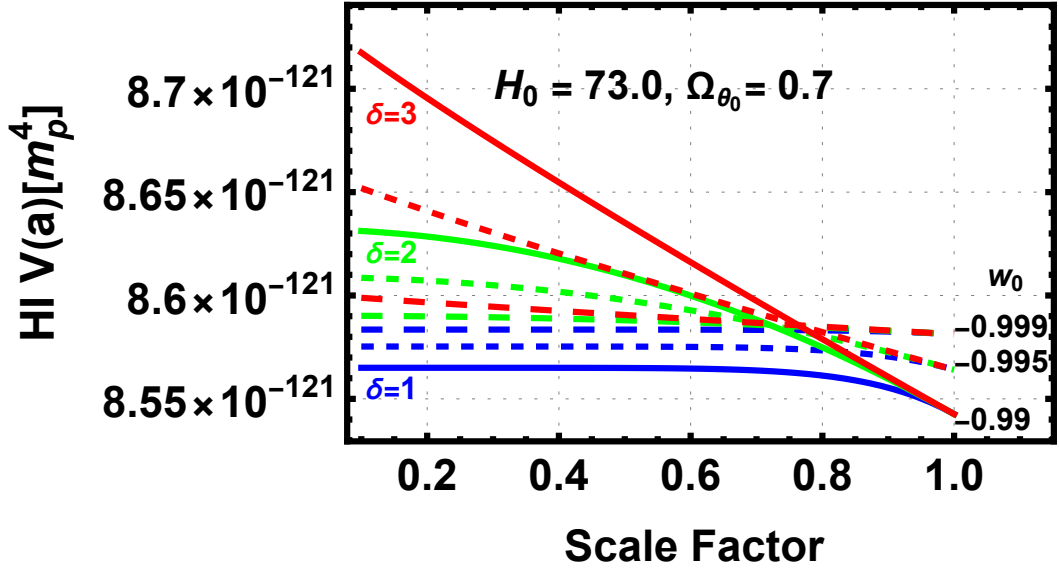}
\caption{The figure shows the evolution of the HI dark energy potential for all cases in this study.}
\label{fig-hipe}
\end{figure}

The evolution of the dark energy potential can now be calculated.  Figure~\ref{fig-hipe} shows the evolution
of the HI dark energy potential for all of the cases.
Figure~\ref{fig-hipe} shows that there is only a small evolution of the potential between scale factors of 0.1 and 1.0,
again consistent with a slow roll.
The maximum evolution is $1.7\%$ for the $\delta=3$ $w_0=-0.99$ case.  The $\delta=1$ evolutions are essentially
constant, mimicking $\Lambda$CDM, until a scale factor of $\approx 0.7$ and then decrease slightly to the $a=1.0$
value for $w_0 = -0.99$ and -0.995.  The $\delta=2$ cases deviate from constant evolution earlier than the $\delta=1$
cases,  The $\delta=3$ cases have almost linear evolution and have the highest values, particularly at small scale
factors.  All of the $w_0=-0.999$ cases have a very flat evolution.  For a given value of $w_0$ the current value of
the potential is the same for all $\delta$ values.  From eqns.~\ref{eq-hirp} and~\ref{eq-mp} the potential at a scale factor 
of one is $3H_0(\Omega_{\theta_0}-\frac{\beta_0^2}{6})$. Equation~\ref{eq-bow} shows that the value of the beta
function at a scale factor of one is $\sqrt{3\Omega_{\theta_0}(w_0+1)}$ making $V_0$ the same for a given $w_0$.

\section{The Hubble Parameter}\label{s-H}
The calculation of the Hubble parameter for the real universe requires the inclusion of both dark energy and matter.
In \citep{cic17} matter is introduced via a differential equation involving the Hubble parameter and the beta function.
Here the Friedmann constraints are the primary tools for deriving the Hubble parameter in a universe with both matter 
and dark energy.  The first Friedmann constraint gives
\begin{equation}\label{eq-hfc1}
3H^2(a)=\rho_{\theta} + \rho_m = \frac{(\kappa\dot{\theta})^2}{2} +M^4((\kappa\theta)^2-(\kappa\delta)^2)^2 +
\frac{\rho_{m_0}}{a^3}.
\end{equation}
Here $\rho_{m_0}$ is the current matter density and $\frac{\rho_{m_0}}{a^3}$ is the mass density as a function of the 
scale factor.  Using equation~\ref{eq-bh}  $\beta H$ is substituted for $\kappa\dot{\theta}$ in equation~\ref{eq-hfc1} to obtain
\begin{equation}\label{eq-fbh}
3H^2(a)(1-\frac{\beta(a)^2}{6})=M^4((\kappa\theta)^2-(\kappa\delta)^2)^2 +\frac{\rho_{m_0}}{a^3}.
\end{equation}
The Hubble Parameter is therefore
\begin{equation}\label{eq-H}
H(a)=\sqrt{\frac{M^4((\kappa\theta)^2-(\kappa\delta)^2)^2 +\frac{\rho_{m_0}}{a^3}}{3\left(1-\frac{\beta(a)^2}{6}\right)}}
\end{equation}

\subsection{The Time Derivative of the Hubble Parameter}\label{ss-hdot}
The second Friedmann constraint in eqns.~\ref{eq-defcon} provides the method for calculating $\dot{H}$.
\begin{equation}\label{eq-hdfc}
\dot{H}=-\left(\frac{\rho_{\theta}(a) +\rho_m(a)+3P(a)}{6} +H^2\right)=-\frac{1}{2}\left((M\kappa)^4\dot{\theta}^2+\frac{\rho_{m_0}}{a^3}\right)
\end{equation}
where $\dot{\phi}=M^2\kappa^2\dot{\theta}$ in units of the reduced Planck mass.

\subsection{The Evolution of the Hubble Parameter}\label{s-eh}
\begin{figure}[H]
\includegraphics[width=14.0 cm]{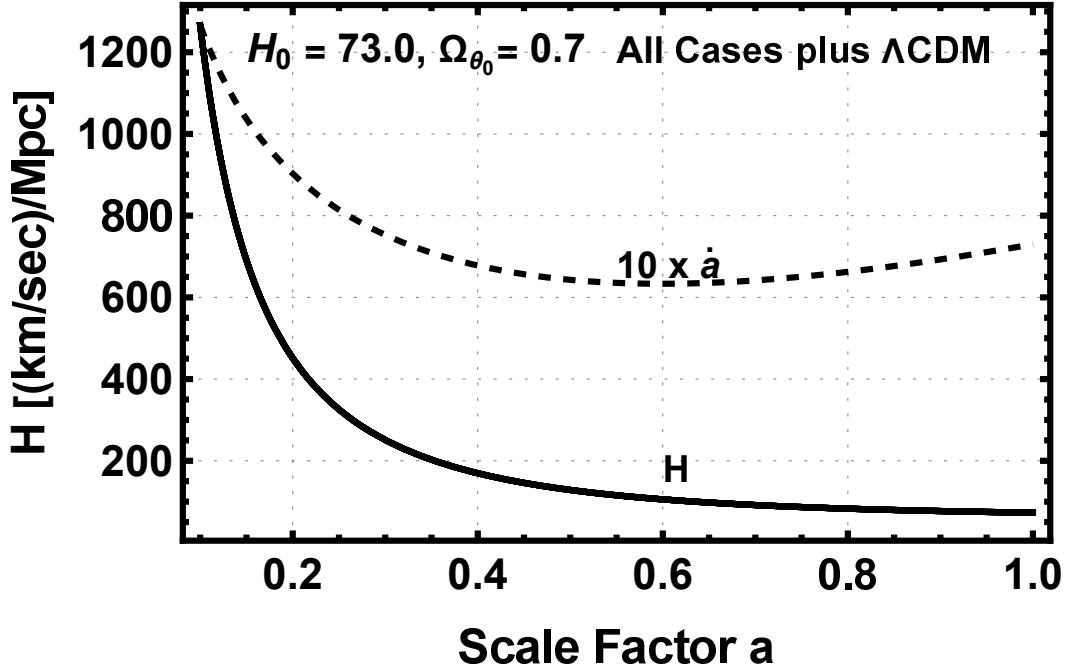}
\caption{The figure shows the evolution of the Hubble parameter for all cases in this study and $\Lambda$CDM.
The evolution of the time derivative of the scale factor is also shown in the dashed line to indicate the onset of the
acceleration of the expansion of the universe.  The scale of its evolution has been magnified by 10 to make it visible 
in the plot.}
\label{fig-eh}
\end{figure}

Figure~\ref{fig-eh} shows the evolution of the Hubble parameter for all values of $\delta$ and $w_0$ plus 
$\Lambda$CDM. At the resolution of the figure all of the tracks overlap each other to the thickness of the line.
The dashed line on figure~\ref{fig-eh} shows the time derivative of the scale factor $\dot{a}$ to show the transition to
acceleration of the expansion of the universe. It occurs at a scale factor of $\approx 0.6$ consistent with current 
observations eg. \citep{dah22}.  The $\dot{a}$ track has been multiplied by ten to remove its overlap with the H
parameter track.  The following section shows the percentage deviation of the HI Hubble parameters from 
$\Lambda$CDM for all of the cases

\subsection{The Percentage Deviation from $\Lambda$CDM}\label{ss-dev}
\begin{figure}[H]
\includegraphics[width=14.0 cm]{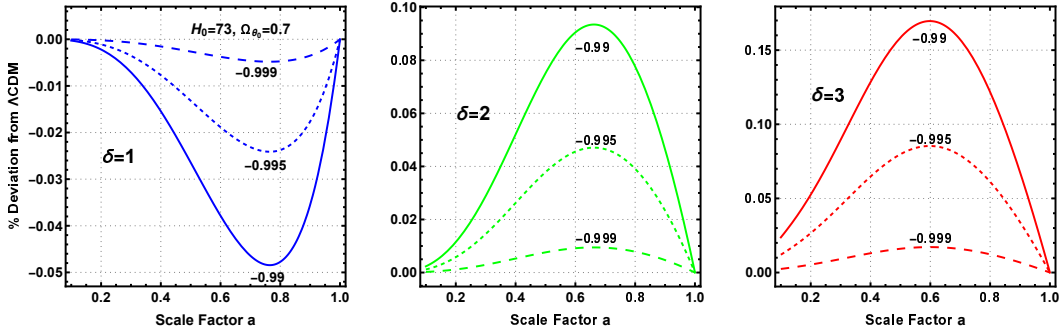}
\caption{The percentage deviation from $\Lambda$CDM for the HI Hubble parameter. The negative numbers 
at the peaks and valleys in each panel are the values of $w_0$.}
\label{fig-perd}
\end{figure}

The fractional deviation of the HI Hubble parameters from $\Lambda$CDM is given by
\begin{equation}\label{eq-dev}
dev=\frac{H_{HI} -H_{\Lambda CDM}}{H_{\Lambda CDM}}
\end{equation}
Figure~\ref{fig-perd} shows the percentage deviation of the HI H parameters from the $\Lambda$CDM H parameter.
The figure readily shows that the percentage deviation of the HI Hubble parameter from $\Lambda$CDM is exceedingly
small.  The highest deviation is $0.17\%$ for the $\delta=3$, $w_0=-0.99$ and the smallest deviation is $0.005\%$ for
the $\delta=1$, $w_0=-0.999$ case.  All of the $\delta=1$ cases have a negative deviation indicating that the HI Hubble
parameter is slightly less than $\Lambda$CDM while the other $\delta$ values have positive deviations with the HI Hubble 
parameter slightly higher than $\Lambda$CDM.  The maximum deviations occur at scale factors between 0.6 for $\delta=3$
and 0.8 for $\delta=1$ where dark energy begins to dominate.  The overall shape of the deviations are reasonable.  
The deviation is zero at $a=1$ since it set by
the $H_0$ boundary condition.  After the peak the deviation drops again as the density becomes matter dominated. 
Currently the deviations of the $w_0=-0.999$ cases are impossible to detect observationally and the highest deviation is
below the detection limit of the proposed near future facilities.  Further discussion of the implications of the HI quintessence
cosmology appears in section~\ref{s-hifc}.

\section{The Scale Factor and Time Derivatives of the Scalar}\label{s-thdev}
The scale factor and time derivatives of the scalar are not observables but are essential for the calculation of the SCP
templates.  The starting point is the derivative of the Lambert W function in equation~\ref{eq-dw}
\begin{equation}\label{eq-dw}
\frac{d W(x)}{dx} = \frac{W(x)}{x(1+W(x))}.
\end{equation}
From this base the derivative of the scalar $(\kappa\theta)$ with respect to the scale factor $a$ is
\begin{equation}\label{eq=thda}
\frac{d\kappa\theta}{da}=\frac{d(\kappa\delta\sqrt{-W(q a^p)})}{da}=\kappa\delta\frac{p\sqrt{-W(q a^p)}}{2a(1+W(q a^p))}.
\end{equation}
The derivative of the scalar with respect to time is then
\begin{equation}\label{eq-dthdt}
\frac{d\kappa\theta}{dt}=\frac{d\kappa\theta}{da}\frac{da}{dt} = \frac{d\kappa\theta}{da}Ha =
\kappa\delta\frac{p\sqrt{-W(q a^p)}}{2(1+W(q a^p))}H.
\end{equation}
Equation~\ref{eq-dthdt} gives the same answer as equation~\ref{eq-bh}.

\begin{figure}[H]
\includegraphics[width=14.0 cm]{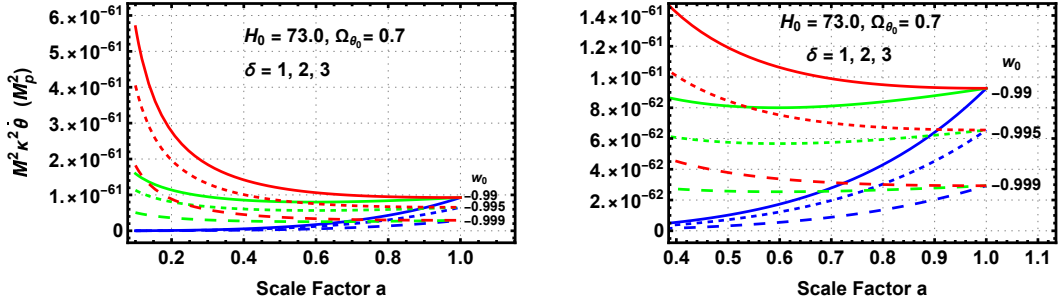}
\caption{The left panel shows the evolution of $M^2 \kappa^2\dot{\theta}$ for all cases and the right panel shows in more detail
the evolution at scale factors between 0.4 and 1.0.}
\label{fig-ktdt}
\end{figure}

Figure~\ref{fig-ktdt} shows the evolution of $\dot{\phi}= M^2 \kappa^2\dot{\theta}$ in units of $m_p^2$ for all of the cases
along with a more detailed plot of the region between a scale factor of 0.4 and 1.0.
The units of $\dot{\theta}$ are $m_p^2$ since $\theta$ has the units of mass and time has units of inverse mass. In
figure~\ref{fig-ktdt} $\kappa^2\dot{\theta}$ is multiplied by $M^2$ to show the value of the time derivative of the 
true scalar $\dot{\phi}$. In the left panel the full evolution of $\dot{\theta}$ is shown for the scale factors between 
0.1 and 1.0 Unlike the scalar the magnitude of $\dot{\theta}$ is decreasing for $\delta=3$ but increasing for $\delta=1$ with
corresponding differences in the second derivative.  The right hand panel shows the evolution between scale factors
of 0.4 and 1.0 in more detail.  Close inspection of the $\delta=2$ and $w_0=-0.99$ track show that it was initially
decreasing but is currently increasing.  This non-monotonic evolution is also present in the dark energy EoS described
later. 

\section{The Dark Energy Density and Pressure}\label{s-dedp}
\begin{figure}[H]
\includegraphics[width=14.0 cm]{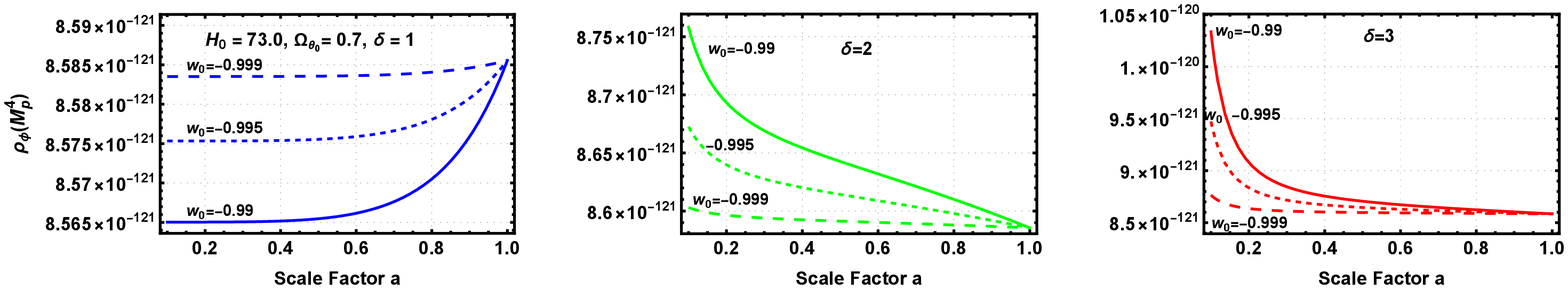}
\caption{The evolution of the dark energy density for all of the cases in this study.}
\label{fig-ded}
\end{figure}

Several cosmological parameters depend on the evolution of the dark energy density and pressure.  Equation~\ref{eq-rhop}
gives the functions for them in terms of the kinetic term $X$ and the dark energy potential.  Figure~\ref{fig-ded} shows the
evolution of the dark energy density for all of the cases.
As usual the $\delta=1$ evolutions have a different character from the other two.  The density evolution for $w_0=-0.999$ is
essentially flat and the highest evolution case, $w_0=-0.99$, only changes by $0.3\%$. For all values of $w_0$ the $\delta=1$
density has a slight rise near a scale factor of one.  For scale factors less than 0.6 the densities are essentially constant
acting like a cosmological constant in the matter dominated epoch.  For $\delta=2$ and 3 the density is monotonically
decreasing with increasing scale factor.  The second derivative of the decrease of density for the $\delta=2$ case  changes
from negative to positive with increasing scale factor similar to the scalar.  The decrease in density for the $\delta=3$ case is
larger than for the other two cases but is still only on the order of $20\%$ for the maximum case.  Unlike the $\delta=2$ case
the second derivative of the evolution is negative at all scale factors.

\begin{figure}[H]
\includegraphics[width=14.0 cm]{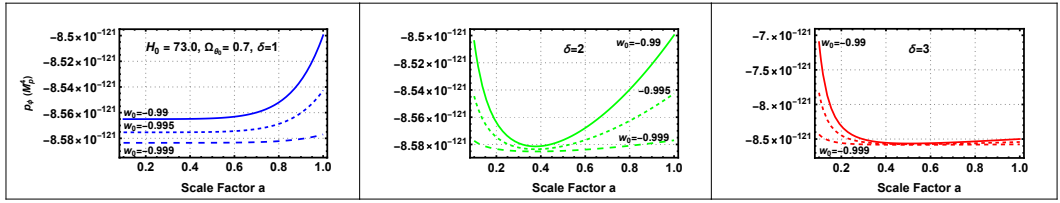}
\caption{The evolution of the dark energy pressure for all of the cases in this study.}
\label{fig-dep}
\end{figure}

Figure~\ref{fig-dep} shows the evolution of the dark energy pressure.
The dark energy pressures have their characteristic negative values and, although more than the density, the absolute 
evolution is relatively small.  The $\delta=1$ and 3 evolutions are monotonically positive and negative respectively but
the $\delta=2$ pressure evolutions have stronger transitions from negative to positive than the density.  As in the dark
energy density the $w_0=-0.999$ evolution is quite flat as would be expected for a $w_0$ value so close to minus one.

\section{The Dark Energy Equation of State}\label{s-w}
\begin{figure}[H]
\includegraphics[width=14.0 cm]{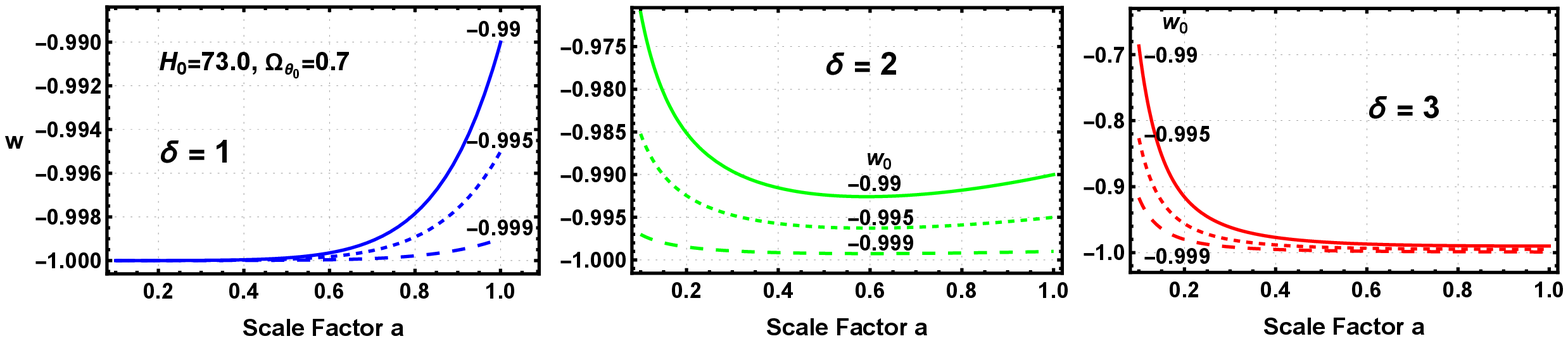}
\caption{The evolution of the dark energy equation of state for all of the cases in this study.}
\label{fig-w}
\end{figure}

By definition the dark energy EoS is the ratio of the dark energy pressure to the dark energy density. Figure~\ref{fig-w}
shows the evolution of $w = \frac{p_{\theta}}{\rho_{\theta}}$.
The $\delta=1$ $w$ evolution is the classic thawing evolution where $w$ is initially near minus one and then thaws to
the less negative values of $w_0$.  The $\delta=3$ is the classic freezing case where $w$ starts at values less negative
than minus one and then freezes toward minus one.  The $\delta=2$ evolution, however, is non-monotonic, starting as
a freezing solution, and then transitioning to a thawing evolution. These evolutions mirror the evolution of the dark energy
pressure in figure~\ref{fig-dep} since the magnitude of the pressure evolution is greater than the evolution of the 
density.
The author does not know of any similar case in the literature but suggests that it may be called the freeze and thaw 
evolution.  Figure~\ref{fig-w} demonstrates the motivation for the simple choices of one, two and three for the $\delta$
values.  The $\delta=3$ cases have late time evolutions very similar to $\Lambda$CDM but significant and observable
deviations at early times.  The $\delta=1$ evolutions of  $w$ are indistinguishable from $\Lambda$CDM at 
early times and only slightly deviant from $\Lambda$CDM at late times due to the purposely chosen $w_0$ 
values very near minus one.  
The $w_0=-0.999$ evolution of $w$ would not be distinguishable from $\Lambda$CDM with current analysis
techniques.  These aspects are discussed more thoroughly in section~\ref{s-hifc} that considers the HI quintessence as a 
candidate for a fiducial dynamical cosmology in the same way that $\Lambda$CDM is a fiducial static cosmology.

\section{The Accuracy of the Cosmology and Dark Energy Potential}\label{s-ac}
At this point the SCP evolutionary templates of all of the cosmological parameters considered in this work are calculated.  
It is appropriate then to consider the accuracy of the cosmology and HI potential as a whole.  The metric for the 
accuracy utilized here is the accuracy of the two Friedmann constraints which contain the Hubble parameter
and its time derivative, the dark energy density and pressure plus the matter density and HI potential.  Other parameters 
such as the dark energy equation of state are functions of the parameters in the Friedmann constraints. The 
first and second Friedmann constraints are given in equation~\ref{eq-friedcs}.  The two constraints are considered separately below.

\subsection{The Accuracy of the First Friedmann Constraint}\label{ss-acf1}
\begin{figure}[H]
\includegraphics[width=14.0 cm]{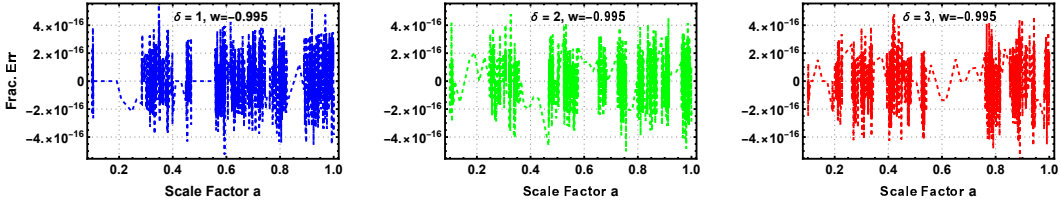}
\caption{The fractional error for the first Friedmann constraint with $w=-0.995$ and $\delta$ = 1, 2, and 3 .}
\label{fig-f1t}
\end{figure}

The left and right sides of the first Friedmann constraint should be equal therefore the accuracy, $fracerr$, is determined by
\begin{equation}\label{eq-f1test}
fracerr=\frac{3H^2-(\rho_{\theta}+\rho_m)}{3H^2}.
\end{equation}
The results for all of the cases are similar in the overall magnitude but of course dissimilar in detail.  The results
for $w_0=-0.995$ and the three values of $\delta$ are shown in figure~\ref{fig-f1t}. 
It is obvious that the first Friedmann constraint is satisfied to better than one part in $10^{16}$. 
This is on the order of the digital accuracy of the Mathematica code used in the calculation.

\subsection{The Accuracy of the Second Friedmann Constraint}\label{ss-acf2}
\begin{figure}[H]
\includegraphics[width=14.0 cm]{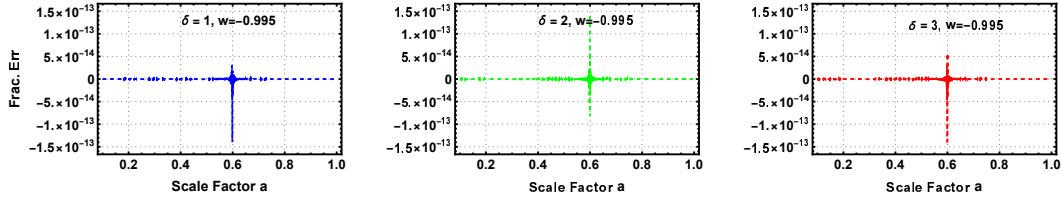}
\caption{The fractional error for the second Friedmann constraint with $w=-0.995$ and $\delta$ = 1, 2, and 3 .
The spikes at $a\approx 0.6$ is due to the denominator passing through zero.}
\label{fig-f2t}
\end{figure}

The second Friedmann constraint explicitly covers more parameters including the time derivative of the Hubble
parameter.  The fractional error for the second Friedmann constraint is given by
\begin{equation}\label{eq-f2test}
fracerr=\frac{3(\dot{H}+H^2)+\frac{(\rho_{\theta} +\rho_m +3P)}{2}}{3(\dot{H} +H^2)}
\end{equation}
Unlike the first Friedmann constraint where all of the terms are positive the second Friedmann constraint has a
mixture of positive and negative terms.  Both the left and right hand term in the numerator of the 
constraint transition between positive and negative values.  The  left hand term is the denominator 
in equation~\ref{eq-f2test} which means that it goes through zero making the fractional error infinite.  The transition occurs 
at a scale factor of approximately 0.6.
Since the calculations in Mathematica are digital true zero rarely occurs but the fractional error does spike at the
transition point.  Figure~\ref{fig-f2t} shows the fractional error for the same cases considered in the first Friedmann
constraint.
The regions away from the spike have similar fractional errors as for the first Friedmann  constraint but build slightly
before the spike.  Even including the spike the second Friedmann constraint is satisfied to a high
accuracy indicating that the SCP evolutionary templates also have a high degree of accuracy, exceeding 
the accuracy of the observations by a high degree.

\section{The HI Quintessence as a Fiducial Dynamical Cosmology}\label{s-hifc}
In most likelihood examinations of cosmological data $\Lambda$CDM is considered the fiducial static cosmology. A
fiducial dynamical cosmology, however, has not been identified.  This may be due in part to the multitude of dynamical
cosmologies and the number of possible dark energy potentials.  This leads to the use of parameterizations and their
incumbent pitfalls as discussed in section~\ref{s-nscpt}. HI quintessence may be a good
candidate for a fiducial dynamical cosmology for comparison with $\Lambda$CDM.  This confronts the 
question of whether dark energy is static or dynamic with the canonical scientific method of comparing 
physics based predictions to the data to measure the likelihood of the predictions.

There are compelling reasons for picking HI quintessence as one of perhaps several fiducial dynamical cosmologies.
A particularly compelling reason is that the HI potential has a natural physical basis since its mathematical form is the
same as the only confirmed isotropic and homogenous field, the Higgs field.  It should be emphasized again here that
the HI scalar is not the Higgs field.  It is just a quintessence scalar field with the mathematical form of the Higgs potential.
Another compelling reason is that, unlike monomial potentials, the HI potential covers a wide range of possible evolutions
by simply varying the value of $\delta$ in the potential.  Section~\ref{s-w} showed that both freezing and thawing
evolutions of the dark energy EoS, $w$, are easily obtained as well as evolutions that transition between freezing and
thawing.  The SCP templates for all of these evolutions are physics based and test real predictions for discriminating 
between static and dynamic dark energy plus determining the nature of a dynamical dark energy.

An additional reason for utilizing HI quintessence as a fiducial cosmology is that it comes arbitrarily close to 
$\Lambda$CDM by varying the constant in the HI potential and adjusting the boundary conditions such as $w_0$
without invoking a cosmological constant. The best example of a $\Lambda$CDM type of evolution in this work is the 
$\delta=1$ and $w_0 = -0.999$ case examined more closely in the next section.

\subsection{A $\Lambda$CDM like dynamical cosmology}\label{ss-dlcdm}
\begin{figure}[H]
\includegraphics[width=14.0 cm]{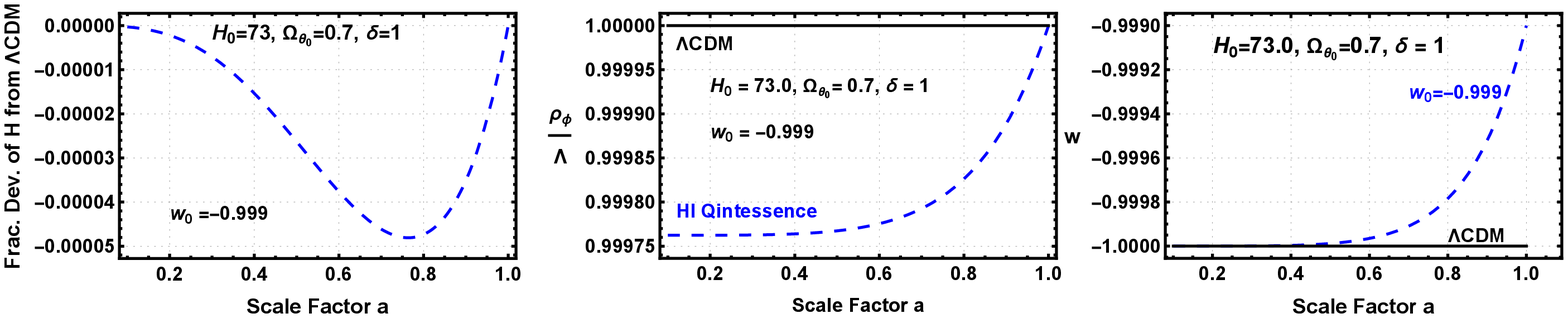}
\caption{The  evolution of the $\delta=1$, $w_0=-0.999$. Hubble parameter, dark energy density and dark energy EoS.
Note that unlike figure~\ref{fig-perd} the fractional deviation of the HI Hubble parameter from $\Lambda$CDM is shown rather
than the percentage.}
\label{fig-hihl}
\end{figure}

Due to the many successes of the $\Lambda$CDM cosmology in matching the observational data the dark energy
EoS $w_0$ values were purposely set close to but not equal to minus one. The $\delta=1$ and $w_0= -0.999$ case
is the closest one to $\Lambda$CDM.  All of the $\delta=1$ cases are thawing which means that the maximum value
of $w$ is $w_0$ and the early time values of $w$ are very close to minus one.  It is this dynamical case of all studied
in this work that has the best chance of having a likelihood close to that of $\Lambda$CDM.  In earlier plots that 
show evolutions for all cases the evolution of this case is often hard to discern since it is much smaller than the 
maximum evolution case in the plots. To better illustrate the $\delta=1$, $w_0=-0.999$ evolutions figure~\ref{fig-hihl}
plots the fractional deviations from $\Lambda$CDM of this case only for the Hubble parameter, the dark energy density 
and the dark energy EoS.
The left panel shows the fractional, not percentage as in figure~\ref{fig-perd}, deviation of the HI Hubble parameter from 
$\Lambda$CDM.  The maximal fractional deviation is only -0.00005 which is below any current or expected near term
detectable limit. The center panel shows the ratio of the dark energy density to the cosmological constant, black line at
1.0, with a maximum deviation of 0.00025 at small scale factors.  The boundary conditions set the deviation at $a=1$
to zero.  The first panel of figure~\ref{fig-ded} indicates that the deviation smaller scale factors than 0.6 is constant
at the $a=0.1$ value.  It is unlikely that the small deviation produces any detectable effects.  The right hand panel shows
the dark energy EoS $w$ which has a maximum deviation from minus one of 0.001.  This also is below current or 
expected near term detection limits.  It is clear that any constraint on the deviation of $w$ can be met by moving $w_0$
closer to minus one, which also lowers the deviations of the other two parameters.  This indicates that it is very difficult
to falsify a dynamical cosmology or to confirm $\Lambda$CDM.  On the other hand a confirmed deviation from the 
$\Lambda$CDM predictions can falsify it but not necessarily confirm a dynamical cosmology.  It would, however, 
produce a higher likelihood for a dynamical cosmology than for $\Lambda$CDM.

\section{Temporal Evolution of Fundamental Constants}\label{s-fc}
Constraints on the temporal and spatial variance of fundamental constants are excellent, but seldom used, discriminators 
between static and dynamic dark energy.  They are also sensitive tests of the validity of the standard model of physics.
Fundamental constants are dimensionless numbers whose values determine the laws of physics.  Primary examples are
the fine structure constant $\alpha$ and the proton to electron mass ratio $\mu$ that are cosmological observables.  Both
are measured by spectroscopic observation of atomic and molecular transitions respectively.  As discussed briefly in the 
introduction the same scalar that produces the late time inflation by interacting with gravity most likely interacts with other 
sectors producing changes in the values of the fundamental constants.  Since the same scalar is determining the value of
the dark energy EoS and the values of the constants there is a relationship that makes the fundamental constants $w$
meters in the universe.  The summation of the interactions of the scalar with  the Quantum  Chromodynamic Scale 
$\Lambda_{QCD}$, the Higgs vacuum expectation value $\nu$ and the Yukawa couplings $h$ produce a net coupling
constant $\zeta_x$ where $x$ can be either $\mu$ or $\alpha$.  In the absence of any knowledge of the coupling it is
assumed to be linear as in equation~\ref{eq-dfc} which can be thought of as the first term of a Taylor series of the real
coupling.
\begin{equation} \label{eq-dfc}
\frac{\Delta x}{x}=\zeta_x (\kappa\theta-\kappa\theta_0)
\end{equation}
Current limits on the temporal variation of the constants are   $\Delta \alpha / \alpha=-(1.3 \pm1.3_{stat} \pm0.4_{sys})
\times 10^{-6}$,  $1\sigma$ \citep{mur21} at $z=1.15$ and  $\Delta \mu / \mu \leq \pm 1.1\times 10^{-7}$ $2\sigma$ 
\citep{bag13,kan15} at $z= 0.89$.  The redshifts for both of these measurements are look back times greater than
half the age of the universe.  The $\alpha$ constraints are from optical spectroscopy of multiple atomic fine structure lines
and the $\mu$ constraints are from radio observations of methanol absorption lines in cold molecular clouds along
the line of sight to a quasar.  In the following the tighter constraint on the temporal variation of $\mu$ is used as an
example.

\begin{figure}[H]
\includegraphics[width=14.0 cm]{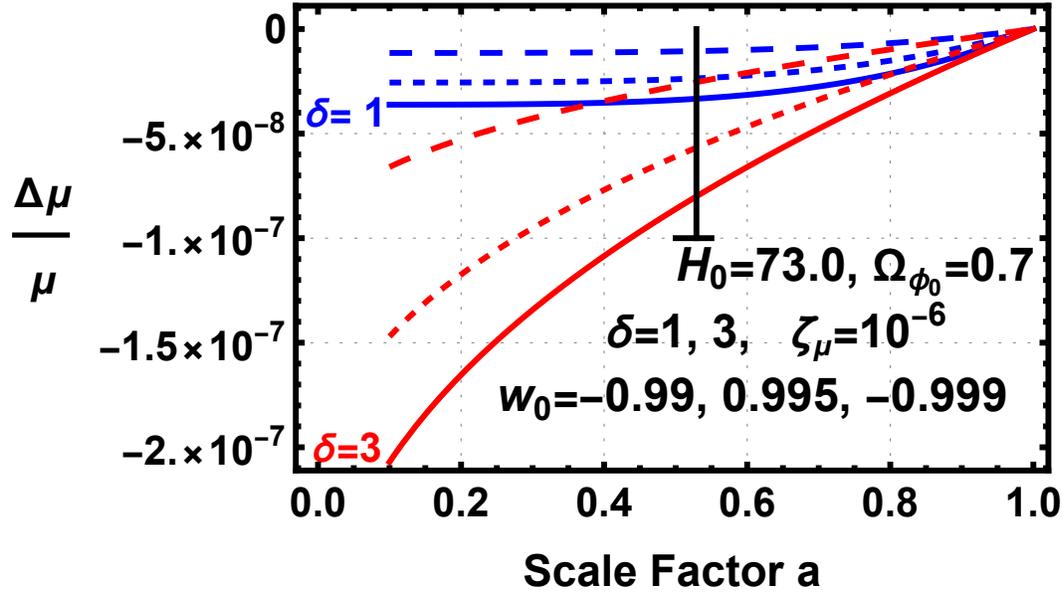}
\caption{The  evolution of $\frac{\Delta \mu}{\mu}$ for $\delta = 1$ and 3 for all three values  of $w_0$. The error bar at a
scale factor of 0.5303 is the $2\sigma$ constraint on the temporal variation of $\mu$.}
\label{fig-mucon}
\end{figure}

Figure~\ref{fig-mucon} shows the evolution of $\frac{\Delta \mu}{\mu}$ for $\delta=1$ and 3 for all three $w_0$ values
and $\zeta_{\mu}=10^{-6}$.
The $\delta=1$ cases represent the least evolution and the $\delta=3$ cases represent the most evolution  
with the $\delta=2$ lying between the two.  All of the cases satisfy the constraint, mainly because of the small 
deviations of $w_0$ from minus one. Only a small tightening of the constraint would start to eliminate some of the
$\delta=3$ cases.  The proposed fiducial case of $\delta=1$ and $w_0=-0.999$ is well within the observational constraint.
Additionally restrictive observational constraints can always be cosmologically accommodated by making $w_0$ closer to
minus one or by lowering the value of the particle physics parameter $\zeta_{\mu}$.

\begin{figure}[H]
\includegraphics[width=14.0 cm]{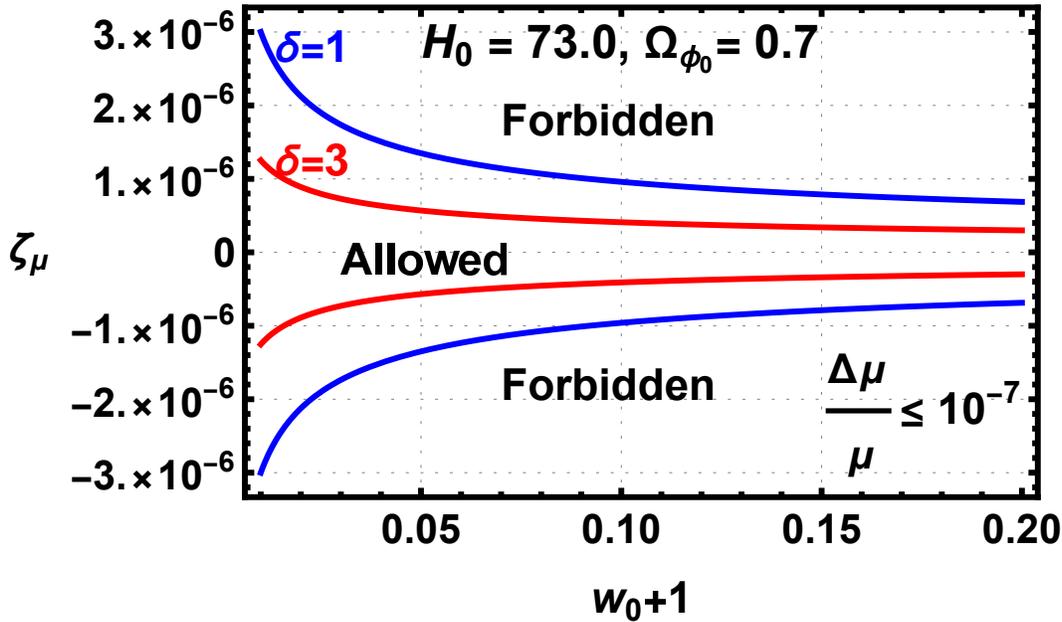}
\caption{The forbidden and allowed regions in the $w_0$ - $\zeta_{\mu}$ plane determined by the constraints on the
temporal deviation of the proton to electron mass ratio $\mu$. Areas inside the two boundary lines for a case 
are allowed while areas outside the two boundary lines are forbidden.}
\label{fig-zwp}
\end{figure}

The last sentence in the above paragraph indicates that a constraint on the temporal variance of either $\mu$ or $\alpha$
is a constraint on a cosmology-particle physics plane defined by $w$ and $\zeta_{\mu}$.  An earlier analysis \citep{thm13} 
determined the relationship between $\zeta_{\mu}$ and $w$ as
\begin{equation}\label{eq-fcp}
\zeta_{\mu} = \frac{\pm\frac{\Delta \mu}{\mu}}{\kappa\theta(a_{ob},w_0,\Omega_{\theta_0})-\kappa\theta_o(1,w_0,\Omega_{\theta_0})}
\end{equation}
where $a_{ob}$ is the scale factor of the observation. Equation~\ref{eq-kto} for $\theta_0$ is the source of the 
$\Omega_{\theta_0}$ term in equation~\ref{eq-fcp}.  Equation~\ref{eq-fcp} defines regions in the $\zeta_{\mu}$-$(w_0+1)$ plane
that satisfy the observational constraint and those that don't.  Figure~\ref{fig-zwp} shows the allowed and forbidden area
for the $\frac{\Delta \mu}{\mu}$ constraint.
The positive and negative tracks for each $\delta$ case are due to the positive and negative values for $\zeta_{\mu}$ in
equation~\ref{eq-fcp}.  The $\delta=3$ more restricted allowed area is due to the greater evolution of 
$\frac{\Delta \mu}{\mu}$.
as shown in figure~\ref{fig-mucon}, that requires a smaller $\zeta_{\mu}$ to meet the constraint than the $\delta=1$ case.  The
$\Lambda$CDM point in the figure is the 0,0, origin where $\zeta_{\mu}=0$ and $(w_0+1)=0$.  A confirmed observation of
$(w_0+1) \ne 0$ with no detected variance of $\mu$ would place a hard limit on the particle physics parameter $\zeta_{\mu}$
but also would require new physics to account for a deviation of $w$ from minus one.

\section{Conclusions}\label{s-con}
This study addresses the question of whether dark energy is static or dynamic by first pointing out that the use of 
parameterizations to represent dynamical cosmologies results in not only erroneous likelihoods but also in
erroneous conclusions about the validity of dynamical cosmologies such as quintessence.  The study then presents a 
methodology for creating accurate analytic templates of the evolution of cosmological parameters and fundamental
constants for the flat quintessence dynamical cosmology.  The methodology utilizes a modified beta function formalism to
determine the evolution of the quintessence scalar as a function of the observable scale factor.  Solutions for the
evolution of parameters and constants that were previously only functions of the unobservable scalar are then translated 
to templates that are functions of the scale factor for direct comparison with the current and expected cosmological
observations.  Recognizing that dynamical cosmologies can have a multitude of dark energy potentials the study
introduced the concept of Specific Cosmology and Potential, SCP, templates to replace the parameterizations with
SCP evolutionary templates based on the physics of the cosmology and dark energy potential.For this reason
the study concentrated on the methodology to produce SCP templates that can embrace a broad range of analytic physics 
based potentials..

To demonstrate the formalism the study then calculated SCP templates of several observable and some necessary
but not observable parameters such as the time derivative of the scalar that appear in the functions of many observable
parameters.  An important aspect of the study is the example quartic polynomial dark energy HI potential.  The modified beta 
function formalism applied to flat quintessence with the HI potential resulted in a scalar that is a simple function of the Lambert 
W function.  This step provided the means to produce accurate analytic SCP templates as a function of the scale factor.

Given the many observational successes of the $\Lambda$CDM static cosmology the study chose boundary conditions
close to $\Lambda$CDM. In particular $w_0$ values close to, but not equal to, minus one were adopted.  This choice
produced a simplification of the beta function formalism where the beta function for quintessence is the negative of the 
logarithmic derivative of a slightly modified dark energy density.    The beta function is then accurately approximated by 
the logarithmic derivative of the potential. Care should be taken in using the formalism for $w$ values significantly 
different than minus one.  Equation~\ref{eq-dpw} shows that the kinetic term $X$ is directly proportional to $(w+1)$.  If $w+1)$ 
becomes too large the approximation can break down and other means must be employed. The SCP templates are calculated 
by imposing the Friedmann constraints on the parameters.  Since the studied epoch included only the matter dominated and
dark energy dominated epochs, radiation is not included in the calculations. This precludes utilization 
of the templates for scale factors smaller than 0.016, such as the CMB dominated epoch, that have significant 
radiation densities.

The polynomial HI potential provided a significantly larger range of evolutions than the often use monomial potentials.  In
particular small changes of the constant term $\delta$ in the potential produced dark energy EoS evolution that were both freezing
and thawing plus evolutions that transitioned from freezing to thawing.  Giving the naturalness of the HI potential and the
large range of evolutions the study suggests that the HI SCP templates become
a fiducial dynamical cosmology in the same way as $\Lambda$CDM is for static cosmologies.  Several of the cases
studied are indistinguishable from $\Lambda$CDM with the accuracy of the present and near future observations even
though their dark energy density arises from a dynamical scalar field rather than a cosmological constant. Given this and
the relative rigidity of the predicted evolutions it appears that  $\Lambda$CDM is easy to falsify but hard to confirm and 
that flat HI quintessence is hard to falsify but easy to confirm if new observations confirm predictions such as a dynamical
dark energy EoS.

The study concluded with an examination of the role of fundamental constants in the discrimination between static and 
dynamical cosmologies.  The scalar in a dynamical dark energy that interacts with gravity will most likely interact with
other sectors which produces temporal variations in the fundamental constants.  To date no confirmed variations of
either $\alpha$ or $\mu$ have been found at the one part in $10^7$ level.  All of the cases in this study predict variations
that are less than the current limits.  Future observations may, however, lower the limit that would make it difficult to meet the 
constraints or find a variation that is consistent with the dynamical predictions.

\section{Appendix 1, Flat HI quintessence abridged templates}
This is an abridged set of evolutionary templates for flat HI quintessence.  The unabridged template set contains significantly
more information including code for implementing the templates.  The templates developed in the main text are gathered here
to provide a convenient listing for community use.  The appendix repeats information provided in the text to gather most of the
relevant material in a single location.

\textbf{Units:} Natural units are utilized with $\hbar$, $c$, and $8\pi G$ set to one.  The units of mass are the reduced Planck mass $m_p$.

\textbf{General constants:}  The constant $\kappa = \frac{1}{m_p}$. In the mass units utilized here $\kappa = 1$ but it is retained to provide
the proper mass units for the templates.

\textbf{Primary variable:} The primary variable is the scale factor $a$.  All templates are functions of the observable scale factor.

\textbf{Special functions:} The Lambert W function $W(x)$ is used extensively in the templates. See \citep{olv10} for a comprehensive
description of the function.

\textbf{The Ratra-Peebles, RP scalar:} The RP scalar is used in all of the templates.  Its functional from is \\
\underline{$\kappa\theta(a) = \kappa\delta\sqrt{-W(q a^p)}=\kappa\delta \sqrt{-W(\chi(a))}$} \\
in terms of the Lambert W function with $q$ and $p$ as constants given below.

\textbf{The Higgs Inspired, HI, dark energy potential:} The dark energy potential is \\
\underline{$V(\kappa\theta) = M^4((\kappa\theta)^2-(\kappa\delta)^2)^2 =M^4((\kappa\theta)^4-2(\kappa\delta)^2(\kappa\delta)^2 +
(\kappa\delta)^4)$} \\
where $M$ is a constant with units of mass in $m_p$. The constant $\delta$ also has units of $m_p$.  Both $V(a)$ and $\kappa\theta(a)$ 
will be repeated below along with the definitions of the constants $q$, $p$ and $M$.

\textbf{Assigned constants:} The HI potential constant $\delta$ is assigned the constants 1.0, 2.0, and 3.0 in this work

\textbf{Changeable cosmological constants:}  These constants are assigned values in this work and appear in the templates, thus
they can be assigned different values according the the desired boundary conditions for the cosmological parameters.  The boundary
conditions are set at the current epoch hence the subscript 0 on their designations.

\underline{$H_0$ the Hubble parameter} \\

\underline{$\Omega_{\theta_0} = \frac{\rho_{\theta_0}}{3H_0^2}$}\\

\underline{$\Omega_{m_0} = \frac{\rho_{m_0}}{3H_0^2}$}\\

\underline{$w_0$ The dark energy equation of state}\\

\textbf{Cosmological parameter templates:}  The cosmological parameter template formats include, where possible, the parameter
first in terms of the RP scalar $(\kappa\theta)$, second the parameter in terms of the Lambert W function,  third its magnitude at
a scale factor of one for $H_0= 73$ $\frac{ km sec^{-1}}{Mpc}$, $w_0 =-0.995$ and $\kappa\delta = 2$ and fourth any associated
constants.
\thicklines
\vspace{0.25cm}
\hrule
\vspace{0.25cm}
\textbf{The Ratra Peebles scalar $\kappa\theta$}\\

\underline{$\kappa\theta(a) =\kappa\delta\sqrt{-W(\chi(a)}$} \\

\underline{$\chi(a)= q a^p$}\\

\underline{ $c=2 (\kappa\delta)^2\ln(\kappa\theta_0)-(\kappa\theta_0)^2$}\\

\underline{$q=-\frac{e^{\frac{c}{(\kappa\delta)^2}}}{(\kappa\delta)^2}$} \\

\underline{$p =\frac{8}{(\kappa\delta)^2}$} \\

\underline{$\kappa\theta_0=-\frac{4-\sqrt{16+12\Omega_{\theta_0}(w_0+1)(\kappa\delta)^2}}{2\sqrt{3\Omega_{\theta_0}(w_0+1)}}$} \\

\underline{$\kappa\theta(1.0) = 0.102202$}
\vspace{0.25cm}
\hrule
\vspace{0.25cm}
\textbf{The beta function $\beta$}\\

\underline{$\beta(a)= -\frac{4\kappa\theta(a)}{(\kappa\theta(a))^2 -(\kappa\delta)^2}$}\\

\underline{$\beta(\chi(a))=\frac{4\sqrt{-W(\chi(a))}}{\kappa\delta(W(\chi(a)) +1)}$}\\

\underline{$\beta(1.0)=0.10247$}\\

\vspace{0.25cm}
\hrule
\vspace{0.25cm}
\textbf{The dark energy potential $V$}\\

\underline{$V(a)=(M\delta)^4((\kappa\theta(a))^2-(\kappa\delta)^2)^2$}\\

\underline{$V(\chi(a))=(M\delta)^4(W(\chi(a)+1)^2$} \\

\underline{$M=\sqrt[4]{\frac{3H_0^2}{((\kappa\theta_0)^2-(\kappa\delta)^2)^2}\left(\Omega_{\theta_0} -\frac{\beta(1)^2}{6}\right)}$}\\

\underline{$V(1.0) =8.56409 \times 10^{-121}$ $m_p^4$}\\
\vspace{0.25cm}
\hrule
\vspace{0.25cm}
\textbf{The Hubble parameter}\\

\underline{$H(a) =\sqrt{\frac{(M \delta)^4((\kappa\theta)^2-(\kappa\delta)^2)^2 +\frac{\rho_{m_0}}{a^3}}
{3\left(1-\frac{\beta(a)^2}{6}\right)}}$}\\

\underline{$H(\chi(a))=\sqrt{\frac{(M\delta)^4(W(\chi(a))+1)^2 +\frac{\rho_{m_0}}{a^3}}
{3\left(1-\frac{\left(\frac{4\sqrt{-W(\chi(a))}}{\kappa\delta(W(\chi(a))+1)}\right)^2}{6}\right)}}$}\\

\underline{$H(1.0)=6.39403 \times 10^{-61}$ $m_p $}\\
\vspace{0.25cm}
\hrule
\vspace{0.25cm}
\textbf{The derivative of the scalar with respect to the scale factor $\frac{d\theta}{da}$}\\

\underline{$\frac{d\theta(a)}{da}=\kappa\delta\frac{p\sqrt{-W(\chi(a))}}{2a(1+W(\chi(a)))}$}\\

\underline{$\kappa\frac{d\theta(1.0)}{da}=0.10247$}\\
\vspace{0.25cm}
\hrule
\vspace{0.25cm}
\textbf{The derivative of the scalar with respect to time $\frac{d\theta}{dt}$}\\

\underline{$\frac{d\theta(a)}{dt}=\dot{\theta}(a)=\frac{d\theta(a)}{da} H(a) a$}\\

\underline{$\frac{d\theta(\chi(a))}{dt}=\kappa\delta\frac{p\sqrt{-W(\chi(a))}}{2(1+W(\chi(a)))}\sqrt{\frac{(M\delta)^4(W(\chi(a))+1)^2 
+\frac{\rho_{m_0}}{a^3}}{3\left(1-\frac{\left(\frac{4\sqrt{-W(\chi(a))}}{\kappa\delta(W(\chi(a))+1)}\right)^2}{6}\right)}}$}\\

\underline{$\frac{d\theta(1.0)}{dt}=6.55193 \times 10^{-62}$ $m_p^2$}\\
\vspace{0.25cm}
\hrule
\vspace{0.25cm}
\textbf{The kinetic term X = $-\frac{\dot{\theta}^2}{2}$}\\

\underline{$X(a) = -\frac{\dot{\theta}(a)^2}{2}=-\frac{1}{2}(\frac{d\theta(a)}{da})^2 H(a)^2 a^2$} \\

\underline{$X(\chi(a))=\left(\kappa\delta\frac{p\sqrt{-W(\chi(a))}}{2(1+W(\chi(a)))}\right)^2\left(\frac{(M\delta)^4(W(\chi(a))+1)^2 
+\frac{\rho_{m_0}}{a^3}}{3\left(1-\frac{\left(\frac{4\sqrt{-W(\chi(a))}}{\kappa\delta(W(\chi(a))+1)}\right)^2}{6}\right)}\right)$}
\vspace{0.25cm}
\hrule
\vspace{0.25cm}
\textbf{The dark energy density}\\

\underline{$\rho_{\theta}(a)=\frac{\dot{\theta}^2(a)}{2}+ (M\delta)^4((\kappa\theta(a))^2-(\kappa\delta)^2)^2$}\\

\underline{$\rho_{\theta}(\chi(a))=-X(\chi(a)) + (M\delta)^4(W(\chi(a)+1)^2$}\\

\underline{$\rho_{\theta}(1)=8.58555 \times 10^{-121}$ $m_p^4$}\\
\vspace{0.25cm}
\hrule
\vspace{0.25cm}
\textbf{The matter density}\\

\underline{$\rho_m(a) = \frac{\rho_{m_0}}{a^3}$}\\

\underline{$\rho_m$ is not a function of $W(\chi(a))$}\\

\underline{$\rho_m(1) = \rho_{m_0} =3.67852 \times 10^{-121}$ $m_p^4$}\\
\vspace{0.25cm}
\hrule
\vspace{0.25cm}
\textbf{The dark energy pressure}\\

\underline{$p_{\theta}(a)=\frac{\dot{\theta}^2(a)}{2}- (M\delta)^4((\kappa\theta(a))^2-(\kappa\delta)^2)^2$}\\

\underline{$p_{\theta}(\chi(a))=-X(\chi(a)) - (M\delta)^4(W(\chi(a)+1)^2$}\\

\underline{$p_{\theta}(1)=-8.54262 \times 10^{-121}$ $m_p^4$}\\
\vspace{0.25cm}
\hrule
\vspace{0.25cm}
\textbf{The dark energy equation of state $w$}\\

\underline{$w(a) = \frac{\frac{\dot{\theta}^2(a)}{2}- (M\delta)^4((\kappa\theta(a))^2-(\kappa\delta)^2)^2}
{\frac{\dot{\theta}^2(a)}{2}+ (M\delta)^4((\kappa\theta(a))^2-(\kappa\delta)^2)^2}$}\\

\vspace{0.15cm}
\underline{$w(\chi(a))=\frac{X(\chi(a)) - (M\delta)^4(W(\chi(a)+1)^2}{X(\chi(a)) + (M\delta)^4(W(\chi(a)+1)^2}$}\\

\underline{$w(1) = -0.995$}\\
\vspace{0.25cm}
\hrule

\section{Referrences}

\end{document}